\PassOptionsToPackage{dvipsnames,table}{xcolor}
\PassOptionsToPackage{shortlabels}{enumitem}
\PassOptionsToPackage{draft}{hyperref}

\documentclass[10pt, sigconf]{acmart}

\renewcommand\footnotetextcopyrightpermission[1]{}
\usepackage[ruled,vlined]{algorithm2e}
\usepackage[english]{babel}
\usepackage[capitalise]{cleveref}
\usepackage{glossaries}
\usepackage{etoolbox}
\usepackage{caption}
\usepackage{subcaption}
\usepackage{amsmath}

\usepackage{enumitem}
\setlist{leftmargin=10pt, 
    itemindent=2pt,
    parsep=2pt,
    topsep=2pt}

\usepackage{pgfplots}
\usepackage{pgfplotstable}
\usepackage{siunitx}
\DeclareSIUnit{\nothing}{\relax}
\usepackage{tikz}
\usetikzlibrary{
    external,
    calc, 
    patterns,
    positioning,
    arrows,
    shadings,
    plotmarks,
    shapes,
    decorations.pathreplacing
}
\usepgfplotslibrary{
    groupplots,
    fillbetween
}
\pgfplotsset{compat=newest}

\setcopyright{acmcopyright}
\newcommand{\carry}{Carry-over}

\newcommand{\eg}{{e.g.,}} 
\newcommand{\ie}{i.e.,} 
\newcommand{\aka}{a.k.a.}

\tikzset{
    legarrow/.style={
        font=\large,
        font=\bfseries,
        inner sep=1pt,
        draw,
        circle, 
        anchor = south
    }
}

\newcommand{\ith}[1]{$#1^{\text{th}}$}

\renewcommand\citet[1]{\citeauthor{#1}~\cite{#1}}

\usepgfplotslibrary{statistics}
\pgfplotsset{compat=newest}
\pgfdeclarelayer{background}
\pgfsetlayers{background,main}

\pgfplotsset{select coords between index/.style 2 args={
    x filter/.code={
        \ifnum\coordindex<#1\def\pgfmathresult{}\fi
        \ifnum\coordindex>#2\def\pgfmathresult{}\fi
    }
}}

\definecolor{colorOrange}{RGB}{237, 109, 0}
\definecolor{colorRed}{RGB}{199, 0, 11}
\definecolor{colorGray}{RGB}{89, 87, 87}
\definecolor{colorGreen}{RGB}{98, 178, 48}
\definecolor{colorBlue}{RGB}{48, 181, 197}

\pgfplotscreateplotcyclelist{MyCyclelist}{%
  {colorRed, thick},
  {colorGray, very thick, densely dotted},
  {colorGreen, very thick, densely dashed},
  {colorOrange, thick, dashdotted},
  {black},
  {colorBlue, mark=x}
}

\pgfplotscreateplotcyclelist{uniformCyclelistFilled}{%
  {colorRed, fill=colorRed!50},
  {colorGray, 
    pattern=crosshatch,
    pattern color = colorGray
  },
  {colorGreen, 
    pattern=mydots,
    pattern color = colorGreen
  }
}

\pgfplotscreateplotcyclelist{uniformCyclelist}{%
  {colorRed},
  {colorGray},
  {colorGreen}
}

\pgfplotscreateplotcyclelist{BenchmarkCycleList}{%
  {colorBlue!80!black, thick, mark=x, mark options={solid}, mark size = 2pt},
  {colorGray, thick, densely dotted, mark=x, mark options={solid}, mark size = 2pt},
  {colorGray, very thick, densely dashed, mark=x, mark options={solid}, mark size = 2pt},
  {colorOrange, thick, thin, mark=x, mark options={solid}, mark size = 2pt},
  {colorOrange, very thick, dashdotted, mark=x, mark options={solid}, mark size = 2pt},
  {black, very thick}
}

\pgfplotscreateplotcyclelist{PrefixCyclelist}{%
  {colorBlue!35!black},
  {colorBlue!35!black, thick, densely dotted},
  {colorBlue!35!black, thick, densely dashed},
  {colorBlue!35!black, thick, dashdotted},
}

\pgfplotscreateplotcyclelist{PrefixCyclelistDual}{%
  {colorBlue!35!black},
  {colorBlue!35!black, thick, densely dotted},
  {colorRed!35!black, thick, densely dashed},
  {colorRed!35!black, thick, dashdotted},
}

\pgfplotsset{compat=1.11,
    /pgfplots/boxplot legend/.style={
        /pgfplots/legend image code/.code={%
        \draw[##1,/tikz/.cd,yshift=-0.25em]
            (0cm,0cm) rectangle (6pt,6pt);},
    },
}

\pgfplotsset{
   boxplot/every median/.style={ultra thick}
}

\pgfplotsset{
	compat=1.11,
	legend image code/.code={
		\draw[mark repeat=2,mark phase=2]
		plot coordinates {
			(0cm,0cm)
			(0.15cm,0cm)        
			(0.3cm,0cm)         
		};%
	}
}

\makeatletter
\pgfplotsset{
    boxplot prepared from table/.code={
        \def\tikz@plot@handler{\pgfplotsplothandlerboxplotprepared}%
        \pgfplotsset{
            /pgfplots/boxplot prepared from table/.cd,
            #1,
        }
    },
    /pgfplots/boxplot prepared from table/.cd,
        table/.code={\pgfplotstablecopy{#1}\to\boxplot@datatable},
        row/.initial=0,
        make style readable from table/.style={
            #1/.code={
                \pgfplotstablegetelem{\pgfkeysvalueof{/pgfplots/boxplot prepared from table/row}}{##1}\of\boxplot@datatable
                \pgfplotsset{boxplot/#1/.expand once={\pgfplotsretval}}
            }
        },
        make style readable from table=lower whisker,
        make style readable from table=upper whisker,
        make style readable from table=lower quartile,
        make style readable from table=upper quartile,
        make style readable from table=median,
        make style readable from table=lower notch,
        make style readable from table=upper notch
}
\makeatother

    \pgfmathdeclarefunction{fpumod}{2}{%
        \pgfmathfloatdivide{#1}{#2}%
        \pgfmathfloatint{\pgfmathresult}%
        \pgfmathfloatmultiply{\pgfmathresult}{#2}%
        \pgfmathfloatsubtract{#1}{\pgfmathresult}%
        \pgfmathfloatifapproxequalrel{\pgfmathresult}{#2}{\def\pgfmathresult{5}}{}%
    }

\newlength{\hatchspread}
\newlength{\hatchthickness}
\newlength{\hatchshift}
\newcommand{\hatchcolor}{}
\tikzset{hatchspread/.code={\setlength{\hatchspread}{#1}},
         hatchthickness/.code={\setlength{\hatchthickness}{#1}},
         hatchshift/.code={\setlength{\hatchshift}{#1}},
         hatchcolor/.code={\renewcommand{\hatchcolor}{#1}}}
\tikzset{hatchspread=3pt,
         hatchthickness=2pt,
         hatchshift=0pt,
         hatchcolor=white}
         
\pgfdeclarepatternformonly[\hatchspread,\hatchthickness,\hatchshift,\hatchcolor]
   {custom north west lines}
   {\pgfqpoint{\dimexpr-2\hatchthickness}{\dimexpr-2\hatchthickness}}
   {\pgfqpoint{\dimexpr\hatchspread+2\hatchthickness}{\dimexpr\hatchspread+2\hatchthickness}}
   {\pgfqpoint{\dimexpr\hatchspread}{\dimexpr\hatchspread}}
   {
    \pgfsetlinewidth{\hatchthickness}
    \pgfpathmoveto{\pgfqpoint{0pt}{\dimexpr\hatchspread+\hatchshift}}
    \pgfpathlineto{\pgfqpoint{\dimexpr\hatchspread+0.15pt+\hatchshift}{-0.15pt}}
    \ifdim \hatchshift > 0pt
      \pgfpathmoveto{\pgfqpoint{0pt}{\hatchshift}}
      \pgfpathlineto{\pgfqpoint{\dimexpr0.15pt+\hatchshift}{-0.15pt}}
    \fi
    \pgfsetstrokecolor{\hatchcolor}
    \pgfusepath{stroke}
   }

\pgfdeclarepatternformonly[\hatchspread,\hatchthickness,\hatchshift,\hatchcolor]
   {custom north east lines}
   {\pgfqpoint{\dimexpr-2\hatchthickness}{\dimexpr-2\hatchthickness}}
   {\pgfqpoint{\dimexpr\hatchspread+2\hatchthickness}{\dimexpr\hatchspread+2\hatchthickness}}
   {\pgfqpoint{\dimexpr\hatchspread}{\dimexpr\hatchspread}}
   {
    \pgfsetlinewidth{\hatchthickness}
    \pgfpathmoveto{\pgfqpoint{\dimexpr\hatchshift-0.15pt}{-0.15pt}}
    \pgfpathlineto{\pgfqpoint{\dimexpr\hatchspread+0.15pt}{\dimexpr\hatchspread-\hatchshift+0.15pt}}
    \ifdim \hatchshift > 0pt
      \pgfpathmoveto{\pgfqpoint{-0.15pt}{\dimexpr\hatchspread-\hatchshift-0.15pt}}
      \pgfpathlineto{\pgfqpoint{\dimexpr\hatchshift+0.15pt}{\dimexpr\hatchspread+0.15pt}}
    \fi
    \pgfsetstrokecolor{\hatchcolor}
    \pgfusepath{stroke}
   }

\pgfdeclarepatternformonly{mydots}
    {\pgfqpoint{-1pt}{-1pt}}
    {\pgfqpoint{1pt}{1pt}}
    {\pgfqpoint{2pt}{2pt}}
{%
  \pgfpathcircle{\pgfqpoint{0pt}{0pt}}{.8pt}%
  \pgfusepath{fill}%
}%

\pgfdeclarepatternformonly
   {mynorthwestlines}
   {\pgfqpoint{-4pt}{-4pt}}
   {\pgfqpoint{8pt}{8pt}}
   {\pgfqpoint{4pt}{4pt}}
   {
    \pgfsetlinewidth{2pt}
    \pgfpathmoveto{\pgfqpoint{0pt}{4pt}}
    \pgfpathlineto{\pgfqpoint{4.15pt}{-0.15pt}}
    \pgfsetstrokecolor{white}
    \pgfusepath{stroke}
   }

\def\heightfigure{2.8cm}

\begin{document}

\copyrightyear{2021} 
\acmYear{2021} 
\setcopyright{acmcopyright}
\acmConference[SEC '21]{ACM/IEEE Symposium on Edge Computing 2021 }{December 14--17, 2021}{San Jose, California, US}

\title{FENXI: Deep-learning Traffic Analytics at the edge}


 \author{Massimo Gallo}
 \affiliation{\institution{Huawei Technologies Co. Ltd} \country{France}}
 \email{massimo.gallo@huawei.com}

 \author{Alessandro Finamore}
 \affiliation{\institution{Huawei Technologies Co. Ltd} \country{France}}
 \email{alessandro.finamore@huawei.com}
 
 \author{Gwendal Simon}
 \orcid{0000-0002-7282-918X}
 \affiliation{\institution{Huawei Technologies Co. Ltd} \country{France}}
 \email{gwendal.simon@huawei.com}
 
 \author{Dario Rossi}
 \affiliation{\institution{Huawei Technologies Co. Ltd} \country{France}}
 \email{dario.rossi@huawei.com}

\renewcommand{\shortauthors}{M. Gallo et al.}

\ccsdesc[500]{Hardware~Analysis and design of emerging devices and systems}
\ccsdesc[500]{Networks~In-network processing}

\newacronym[longplural={Frames per Second}]{fpsLabel}{FPS}{Frame per Second}
\newacronym{dl}{DL}{Deep Learning}
\newacronym{ml}{ML}{Machine Learning}
\newacronym{tpu}{TPU}{Tensor Processing Unit}
\newacronym{cnn}{CNN}{Convolutional Neural Network}
\newacronym{ai}{AI}{Artificial Intelligence}
\newacronym{gpu}{GPU}{Graphic Processing Unit}
\newacronym{nic}{NIC}{Network Interface Card}
\newacronym{cpu}{CPU}{Central Processing Unit}
\newacronym{numa}{NUMA}{Non-Uniform Memory Access}
\newacronym{dma}{DMA}{Direct Memory Access}
\newacronym{pcie}{PCIe}{Peripheral Component Interconnect Express}
\newacronym{iad}{IAT}{Inter-Arrival Time}
\newacronym{fix}{FENXI}{Fast In-Network Analytics}
\newacronym{qoe}{QoE}{Quality of Experience}
\newacronym{simd}{SIMD}{Single Instruction Multiple Data}
\newacronym{int}{INT}{In-band Network Telemetry}
\newacronym{asic}{ASIC}{Application-specific integrated circuit}
\newacronym{lstm}{LSTM}{Long Short-Term Memory}
\newacronym{isp}{ISP}{Internet Service Provider}
\newacronym{mawi}{MAWI}{Measurement and Analysis on the WIDE Internet}
\newacronym{pop}{POP}{Point-of-Presence}
\newacronym{cdf}{CDF}{Cumulative Distribution Function}
\newacronym{dpi}{DPI}{Deep Packet Inspection}
\newacronym{sni}{SNI}{Server Name Identification}
\newacronym{rest}{REST}{Representational State Transfer}
\newacronym{rpc}{RPC}{Remote Procedure Call}
\newacronym{mpu}{MPU}{Main Processing Unit}
\newacronym{rtt}{RTT}{Round-Trip Time}
\newacronym{wan}{WAN}{Wide Area Network}
\newacronym{lru}{LRU}{Least Recently Used}
\newacronym{dpdk}{DPDK}{Data Plane Development Kit}
\newacronym{rss}{RSS}{Receive Side Scaling}
\newacronym{pisa}{PISA}{Protocol Independent Switch Architecture}
\newacronym{ar}{AR}{Augmented Reality}
\newacronym{vr}{VR}{Virtual Reality}

\begin{abstract}
Live traffic analysis at the first aggregation point in the ISP network enables the implementation of complex traffic engineering policies but is limited by the scarce processing capabilities, especially for \gls{dl} based analytics. The introduction of specialized hardware accelerators \ie{} \gls{tpu}, offers the opportunity to enhance processing capabilities of network devices at the edge. Yet, to date, no packet processing pipeline is capable of offering DL-based analysis capabilities in the data-plane, without interfering with network operations.

In this paper, we present \acrshort{fix}, a system to run complex analytics by leveraging \gls{tpu}. The design of \acrshort{fix} decouples forwarding operations and traffic analytics which operates at different granularities \ie{} packet and flow levels. 
We conceive two independent modules that asynchronously communicate to exchange network data and analytics results, and design data structures to extract flow level statistics without impacting per-packet processing. 
We prototyped and evaluated \acrshort{fix} on general-purpose servers considering both both adversarial and realistic network conditions. Our analysis shows that \acrshort{fix} can sustains \SI{100}{\giga bps} line rate traffic processing requiring only limited resources, while also dynamically adapting to variable network conditions.

\end{abstract}

\keywords{Traffic Measurement, Deep Learning, Real-Time}

\maketitle

\glsresetall
\section{Introduction}

In the last decade, \gls{dl} has become a fundamental analytics technique in some fields of computer science such as computer vision and natural language processing. 
\gls{dl}-based analytics are rapidly gaining momentum in the network community too, with the promise of enabling complex \gls{ai}-based traffic engineering. 
Proposals that use \gls{dl} models to improve traffic engineering include application identification~\cite{PachecoEGBA19,AcetoCMP19}, analytics that enables traffic differentiation~\cite{LiNCGM19,XuTMZWLY18}, malware and attack detection~\cite{DavidN15,MarinCC18} used by firewall applications, and anomaly detection~\cite{troubleshooting} by troubleshooting tools.
However executing pre-trained \gls{dl} models (\aka{} inference) is still a complex operation, which requires significant processing capabilities that are not commonly found in the network edges. Indeed, the research community has concentrated efforts in designing systems that delegate inference to external devices, typically hosted in a cloud environment~\cite{MestresRCBASMMB17,clipper}. 

We consider traffic monitoring operated at the network edge, 
typically in the first aggregation point after the so-called ``last-mile'' in broadband networks. Despite the increasing interest in both industry and research communities, the adoption of \gls{dl} has yet to transform network management and traffic monitoring in this scenario. On the one hand, current approaches which offload \gls{dl} inference to the cloud, alleviate the demand for increasing physical resources at the edge~\cite{convnet}. On the other hand, cloud offloading is both ineffective for latency-sensitive use-cases (\eg{} when decisions should be taken within a few \gls{rtt} of the flow life cycle) and unable to deal with the constant increase of broadband link capacity (\eg{} resulting in overwhelming control traffic to the cloud services) as pointed out in \cite{XiongZ19}.

In this paper we study the design of an inference system that leverages low cost and low power consumption \emph{hardware accelerators}~\cite{HPECsurvey19} such as edge \gls{tpu}~\cite{ascend310,coral}, which will be included in the next-generation network cards~\cite{mellanox}. We design, implement, and extensively benchmark \acrshort{fix},
a system for \gls{dl} traffic analytics at line rate. With \acrshort{fix}, we revisit packet processing pipelines design from the \gls{nic} to the storage of the inference result, contributing to the field in two ways: ($i$) we report a deep dive into a challenging system with multiple contrasting objectives, and ($ii$) we propose practical solutions to every element of the system.

\textbf{Contribution \#1}: By implementing and testing \acrshort{fix} under realistic and extreme conditions, we reveal the complexity of \gls{dl} analytics in edge network scenarios.

\begin{itemize}
    \item We benchmark \gls{dl} hardware accelerators: a \gls{gpu}, a \gls{tpu}, and a multi-core CPU. We used a state-of-the-art \gls{dl} model designed for traffic analytics~\cite{AcetoCMP19} to test the performance of the accelerator. The benchmark report in \Cref{sec:rqrmnt} highlights the trade-off between inference speed (to sustain throughput), delay (to meet application requirements), and energy consumption.
    \item We examine the characteristics of real network traffic to identify system requirements. Our analysis emphasizes two requirements: $(i)$ high-throughput operations considering the number of packets per second (since information is extracted from individual packets) and the number of flows per second (since analytics apply on flows). $(ii)$  low-delay operations since the delay in inferring the analytics is not only a matter of processing a data stream but also an application requirement that comes from the traffic characteristics and the analytics usage, \eg{} post-mortem analytics are not useful for traffic management.
\end{itemize}

\noindent\textbf{Contribution \#2}: We introduce \acrshort{fix} architecture, together with a thorough analysis of the potential bottlenecks, exposing pitfalls that should be avoided in the design. 
\begin{itemize}
	\item We present in \Cref{sec:feature} the architecture of the Flow manager, which is responsible for \emph{extracting features} from the flows of packets without interfering with the traffic forwarding. 
	We corroborate the design choices with a set of micro-benchmark to illustrate the performance of each individual building block.
	\item We present in \Cref{sec:batching} a \emph{dynamic batching system}, which addresses the threefold requirements of sustaining throughput, maintaining low delay, and reducing energy usage. The concept of grouping multiple data in batches is fundamental in inference serving systems but introduces latency. Some researchers have regarded batching as a not viable solution for latency-sensitive applications~\cite{SanvitoSB18,JouppiYPPABBBBB17}. We study this supposed mismatch and improve over state-of-the-art solutions~\cite{JinGXSNLQW19} by introducing a mechanism to reduce processing and energy usage waste.
	\item We introduce in \Cref{sec:caching} a dedicated \emph{caching system}, which we specifically designed for analytics of network flows. Previous papers have shown that caching for packet inference suffers from header entropy~\cite{swamy2020taurus}: we address this problem by designing approximate caching policies that suit packet series in flows and evaluate them through a set of micro-benchmarks.
	\item We finally evaluate in \Cref{sec:am_bench} the performance of the whole \acrshort{fix} system under scenarios that are the most challenging with respect to our objective of implementing analytics in the data path.
\end{itemize}

\section{Context and Requirements}\label{sec:rqrmnt}

We describe in this section the parameters and constraints that interplay in the design of traffic analytics pipelines. Whereas the main principles of \acrshort{fix} broadly apply to multiple \gls{dl} analytics models and network scenarios, we further introduce a specific case study with the aim of clarifying the challenges and providing tangible numerical objectives. We first describe the regarded case study, and then, we emphasize the main operational points of the system.

\def\relativepower{30}
\def\relativerate{50000}
\def\relativeratek{50}

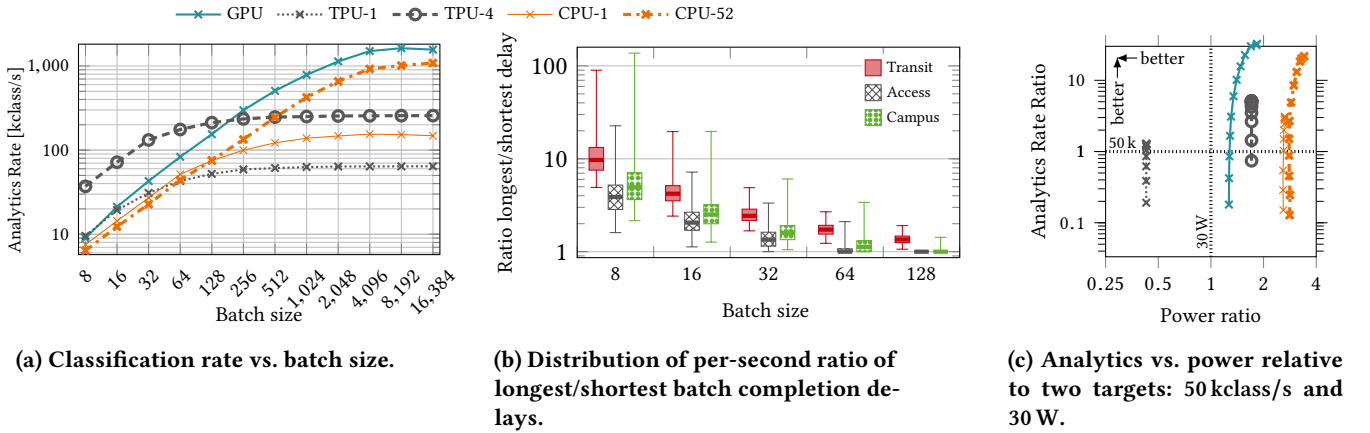
\begin{figure*}[ht!]
\centering
\def\ratioFig{0.30} 
\def\ratioQuad{0.25}
\def\xrat{0.27} 
\def\yrat{1} 
\def\xlab{-0.2}

\pgfplotsset{
	compat=1.11,
	legend image code/.code={
		\draw[mark repeat=2,mark phase=2]
		plot coordinates {
			(0cm,0cm)
			(0.3cm,0cm)        
			(0.6cm,0cm)         
		};%
	}
}

\pgfplotstableread[col sep = tab]
    {Data/inference_gpu.log}
    \tableGPU

\pgfplotstableread[col sep = tab]
    {Data/inference_asc-1c-2t.log}
    \tableTPU

\pgfplotstableread[col sep = tab]
    {Data/inference_asc-4c-2t.log}
    \tableTPUtwo

\pgfplotstableread[col sep = tab]
    {Data/inference_cpu-1c.log}
    \tableCPU

\pgfplotstableread[col sep = tab]
    {Data/inference_cpu-52c.log}
    \tableCPUtwo

\subcaptionbox{Classification rate vs. batch size.\label{fig:hw_classrate}}
	[\ratioFig\textwidth] 
	{\begin{tikzpicture}
    \begin{axis}[
        ymode=log, log ticks with fixed point,
        scaled y ticks=false,
        xmode=log, log ticks with fixed point,
        scaled x ticks=false,
        ylabel near ticks,
        xtick=data,
        /pgf/number format/.cd,fixed,precision=0,
        xticklabel style={
            rotate=45,
            scaled ticks=false
        },
        xticklabel=\pgfmathparse{exp(\tick)}\pgfmathprintnumber{\pgfmathresult},
        ylabel={Analytics Rate [kclass/s]},
        scale only axis,
        width = \xrat*\textwidth, 
        height = \yrat*\heightfigure,
        xlabel = Batch size,
        ylabel style={
            align=center},
        yticklabel style={
            font = \scriptsize
        },
        xticklabel style={
            font = \footnotesize
        },
        label style = {
            font=\footnotesize
        },
        ylabel style = {
            font = \scriptsize,
            inner sep = 1pt,
            at={(axis description cs:-0.14,0.45)}
        },
        x label style={
            font=\footnotesize,
            at={(axis description cs:0.5,\xlab)},
            anchor=north},
        legend style={
            font=\scriptsize,
            at={(0.25, 1.05)},
            anchor = south west,
            draw = none,
            fill = none,
            inner sep=2pt,
            /tikz/every even column/.append style={
                column sep=3pt}
               },
        legend columns=5,
        legend cell align={left},
        ymajorgrids,
        xmajorgrids,
        yminorgrids,
        enlargelimits=0.02,
        cycle list name=BenchmarkCycleList
    ]

    \addplot+ [] table [
                x index = 0, 
                y expr = \thisrowno{1}/1000
            ] {\tableGPU};
    \addlegendentry{GPU}
    
    \addplot+ [] table [
                x index = 0, 
                y expr = \thisrowno{1}/1000
            ] {\tableTPU};
    \addlegendentry{TPU-1}
    
    \addplot+ [mark = o] table [
                x index = 0, 
                y expr = \thisrowno{1}/1000
            ] {\tableTPUtwo};
    \addlegendentry{TPU-4}
    
    \addplot+ [] table [
                x index = 0, 
                y expr = \thisrowno{1}/1000
            ] {\tableCPU};
    \addlegendentry{CPU-1}
    
    \addplot+ [] table [
                x index = 0, 
                y expr = \thisrowno{1}/1000
            ] {\tableCPUtwo};
    \addlegendentry{CPU-52}
    \end{axis}
\end{tikzpicture}}
\hspace{-0.5cm}
\hfill%
\subcaptionbox{Distribution of per-second ratio of longest/shortest batch completion delays.
    \label{fig:bursts}}
    [\ratioFig\textwidth]
    {\def\rat{0.12}

\begin{tikzpicture}
    \begin{axis}[
        scale only axis,
        boxplot/draw direction=y,
        boxplot = {
            draw position={1/4 + floor(\plotnumofactualtype/3) + 1/4*mod(\plotnumofactualtype,3)},
            box extend=0.19,
        },
        ymode=log, log ticks with fixed point,
        scaled y ticks=false,
        ymin=1,
        xmin=0, xmax=5,
        width = \xrat*\textwidth, 
        height = \yrat*\heightfigure,
        xlabel = Batch size \vphantom{p\ith{10}},
        ylabel = Ratio longest/shortest delay,
        ylabel style = {
            align = center,
            font=\footnotesize,
            at={(axis description cs:-0.15,.5)},
            inner sep=1pt
            },
        xlabel style = {
            font=\footnotesize},
        x=\rat*\columnwidth, 
        xticklabels={
            {0},
            {8},%
            {16},%
            {32},%
            {64},%
            {128}},
        x tick label style={
            text width=0.05*\columnwidth,
            align=center,
            font = \footnotesize
        },
        xtick distance=1,
        x tick label as interval,
        legend style={
            font = \tiny,
            draw = none,
            fill = white,
            inner sep = 1pt,
        },
        legend pos = north east, 
        legend columns=1,
        legend cell align={left},
        boxplot legend,
        ymajorgrids,
        enlarge y limits=0.03,
        cycle list name=uniformCyclelistFilled
    ]
    \pgfplotstableread[col sep = comma]
    {Data/burst.csv}
    \mytable

    \pgfplotstablegetrowsof{\mytable}
    \pgfmathsetmacro\numberofrows{\pgfplotsretval-1}

    \pgfplotsinvokeforeach{0,...,\numberofrows}{
        \addplot+[
            boxplot prepared from table={
                table=\mytable,
                row =#1,
                lower whisker=min_transit,
                upper whisker=max_transit,
                lower quartile=lw_transit,
                upper quartile=uw_transit,
                median=med_transit
                }, 
            boxplot prepared]
        coordinates {};
        \addplot+[
            boxplot prepared from table={
                table=\mytable,
                row =#1,
                lower whisker=min_access,
                upper whisker=max_access,
                lower quartile=lw_access,
                upper quartile=uw_access,
                median=med_access
                }, 
            boxplot prepared]
        coordinates {};
        \addplot+[
            boxplot prepared from table={
                table=\mytable,
                row =#1,
                lower whisker=min_campus,
                upper whisker=max_campus,
                lower quartile=lw_campus,
                upper quartile=uw_campus,
                median=med_campus
                }, 
            boxplot prepared]
        coordinates {};
    }
    \legend{Transit, Access, Campus}
    \end{axis}
\end{tikzpicture}}%
\hfill%
\subcaptionbox{Analytics vs. power relative to two targets:  \SI{\relativeratek}{\kilo class/s} and \SI{\relativepower}{\watt}.\label{fig:hw_quadrant}}
	[\ratioQuad\textwidth] 
	{\begin{tikzpicture}

\def\maxyaxis{30}
\def\maxxaxis{4}

\begin{loglogaxis}[
    tick align=outside,
    scale only axis,
    width = \heightfigure, 
    height = \heightfigure,
    xlabel={Power ratio},
    log ticks with fixed point,
    xtick = {0.25, 0.5, 2, 4},
    ylabel={Analytics Rate Ratio},
    xmin=1/\maxxaxis, xmax=\maxxaxis,
    ymin=1/\maxyaxis, ymax=\maxyaxis,
    ytick style={color=black},
    yticklabel style={
            font = \scriptsize
        },
    xticklabel style={
            font = \footnotesize
        },
    extra x ticks={1},
    extra y ticks={1},
    extra tick style={
        grid=major,
        major grid style={
            densely dotted, 
            very thick,
            black}},
    x label style={
        font=\footnotesize,
        at={(axis description cs:0.5,\xlab)},
        anchor=north},
    y label style={
        font=\footnotesize,
        at={(axis description cs:-0.40,0.5)},
        anchor=north},
    cycle list name=BenchmarkCycleList,
    clip = false
]


\node[draw = none,
    inner sep = 1pt, 
    anchor = north west,
    font = \scriptsize]
    at (rel axis cs: 0.15, 0.99)
    (legbetterx)
    {better};
\draw [-latex]
    (legbetterx.west) to
    ([xshift=-8pt] legbetterx.west);
    
\node[draw = none,
    inner sep = 1pt, 
    anchor = north east,
    font = \scriptsize,
    rotate = 90]
    at (rel axis cs: 0.01, 0.85)
    (legbettery)
    {better};
\draw [-latex]
    (legbettery.east) to
    ([yshift=8pt] legbettery.east);

\node[draw=none,
    inner sep=1pt,
    anchor=south west,
    font=\tiny,
    rotate=90]
    at (rel axis cs: 0.50, 0.02)
    {\SI{\relativepower}{\watt}};
\node[draw=none,
    inner sep=1pt,
    anchor=south west,
    font=\tiny,
    ]
    at (rel axis cs: 0.0, 0.5)
    {\SI{\relativeratek}{\kilo\nothing}};

    

\addplot+ [] table [
        x expr = \thisrowno{2}/\relativepower, 
        y expr = \thisrowno{1}/\relativerate]
            {\tableGPU};

\addplot+ [] table [
        x expr = 12.8/\relativepower, 
        y expr = \thisrowno{1}/\relativerate]
            {\tableTPU};
    
\addplot+ [mark = o] table [
        x expr = 4*12.8/\relativepower, 
        y expr = \thisrowno{1}/\relativerate,
        ]
            {\tableTPUtwo};
\addplot+ [] table [
        x expr = \thisrowno{2}/\relativepower, 
        y expr = \thisrowno{1}/\relativerate]
            {\tableCPU};
\addplot+ [] table [
        x expr = \thisrowno{2}/\relativepower, 
        y expr = \thisrowno{1}/\relativerate]
            {\tableCPUtwo};

\end{loglogaxis}

\end{tikzpicture}}%
\caption{Throughput requirements, boxplots show \ith{99}, \ith{75}, \ith{25}, and $1^{\text{st}}$ percentiles.}
\label{fig:hw}
\end{figure*}

\subsection{Case Study}

Instead of an in-breadth analysis of multiple use-cases, we opt for an in-depth analysis of one use-case. We focus on \emph{application identification} as a classic example of flow-level traffic analytics. The identification of the specific application related to a flow is a strategic network management operation. For this task, the inference is triggered after having observed a sufficient (but small) number of packets for each flow~\cite{teixeira06ccr,crotti07ccr}. Traffic classification has received growing attention from researchers in the \gls{dl} community~\cite{AcetoCMP19}. 
We chose traffic classification among other analytics due to its challenging requirements: 
($i$) all packets need to be processed and recomposed into flows; ($ii$) the first packets of each flow is used for classification; ($iii$) the classification need to be as fast as possible (\ie{} the identified label is useless if determined after the flow is completed). In comparison, other traffic analytics have less stringent requirements.

We operate traffic classification via a 1D-\gls{cnn} model, which size (about \SI{100}{\kilo\nothing} weights) is smaller than typical 2D \gls{cnn} models used for image processing, but is significantly larger than the toy-case models used in the related system work~\cite{swamy2020taurus,abs-2009-02353}. 
The model is equivalent to the one used in \cite{pandorabox} trained with over \num{200} applications labels, which is about ten (four) times the typical (maximum) number of classes considered in the literature~\cite{AcetoCMP19}.

Flow classification requires the extraction of IP packets for the analysis of network flows, as identified by the \num{5}-tuple at the network layer (IP addresses and ports of both source and destination plus protocol). For this specific analytics, \acrshort{fix} triggers, at the \ith{K} packet of each flow, the DL inference based on the size of packets. For other analytics, \acrshort{fix} is capable of using other flow information and triggers. By default in this paper, we use $K$ equal to \num{10}, which corresponds to a trade-off between the amount of information to analyze and the delay to process the analytics~\cite{AcetoCMP19}. Formally we extract from each flow a \emph{series} $S$ containing $K$ information, which are extracted from the $K$ first packets $s_1, \dots, s_K$ of the flow. We denote by $\mathcal S_K$ the set of series of size $K$. In the case of traffic classification, the information $s_i$ extracted from the \ith{i} packet of a flow consists in packet length and direction. Other analytics may require information such as \gls{iad}~\cite{AcetoCMP19} or transport-level flags.

We benchmark the performance of the \gls{dl} model on the hardware accelerators. We ran preliminary experiments on servers equipped with Intel Xeon Platinum 8164 CPUs @ 2.00GHz (L1/L2/L3 caches 32 data+32 instruction/1024/36608 kB) and 100 Gbps Mellanox MCX515A-CCAT ConnectX-5. As \gls{dl} hardware accelerator, we used either a Huawei Atlas 300I:3010 TPU inference card (equipped with \num{4}$\times$ Ascend 310 chips) or an Nvidia V100 GPU. To provide a fair comparison between \gls{tpu} and \gls{gpu}, we did not port the model to the native Huawei Mind Studio stack; we rather cross-compiled the original TensorFlow model for the Huawei Atlas engine. 

\acrshort{fix} targets network devices that operate at the edge of the Internet and contribute to network management operations. To analyze the requirements of such devices, we studied two representative datasets: $(i)$ a gateway that connects a \textit{Campus} network to the \gls{wan} Internet, this device is representative of enterprise private networks; and $(ii)$ a \gls{pop} router that connects a residential \textit{Access} network to the Internet. Since both datasets are private, for the sake of transparency, we also used a third dataset (called \textit{transit}), which comes from the public MAWI project.\footnote{\url{http://mawi.wide.ad.jp/} -- extracted 2020-02-12, 2020-03-04, 2020-03-25, 2020-04-08, 2020-05-27, 2020-06-03, 2020-06-10} 
For each dataset, we extracted relevant traffic characteristics by identifying all flows by means of the classic IP \num{5}-tuple, from which we extracted packet time-stamps and lengths that are reported in \Cref{tab:dataset} which is used to motivate system requirements. 


\begin{table}[t!]
    \centering
    \scriptsize
    \setlength{\tabcolsep}{4pt}
    \begin{tabular}{@{}cccccccc}
        \toprule
         & \multicolumn{4}{c}{\textbf{Traffic characteristics}} & \multicolumn{3}{c}{\textbf{ Rate at 100 {Gbps}}}
         \\ \cmidrule(l){2-5} \cmidrule(l){6-8}
         \textbf{dataset} & \textbf{vol.} & \textbf{\#{}pkts} & \textbf{\#{}flows} & \textbf{\#{}series}  & \textbf{packet} & \textbf{flow} & \textbf{series}
         \\ & [\SI{}{\giga B]} &  [\SI{}{\mega\nothing]} & 
         [\SI{}{\kilo\nothing]} & [\SI{}{\kilo\nothing]} & 
         [\SI{}{\mega pps}]& [\SI{}{\kilo flows/s}] & [\SI{}{\kilo class/s}]
         \\ \midrule 
access & \num{765} & \num{858}
    & \num{3963} & \num{2481}
    & \num{14.0} & \num{64.7} & \num{40.5} \\
transit & \num{870} & \num{923}
    & \num{2476} & \num{1968}
    & \num{13.2} & \num{35.5} & \num{28.3} \\
campus & \num{483} & \num{516}
    & \num{2700} & \num{1718}
    & \num{13.3} & \num{69.8} & \num{44.4} \\
\midrule 
average & \num{706} & \num{765} 
    & \num{3046} & \num{2055} 
    & \num{13.4} & \num{56.6} & \num{37.3} \\
         \bottomrule
    \end{tabular}
    \caption{Traffic characteristics in real datasets.}
    \label{tab:dataset}
\end{table}

\begin{figure*}[ht!]
\def\ratio{0.32}
\def\yrat{1}
\def\xrat{0.23}
\def\ratioFig{0.30} 
\def\ratioQuad{0.25}
\def\xrat{0.27} 
\def\yrat{1} 
\def\xlab{-0.2}
\pgfplotsset{
	compat=1.11,
	legend image code/.code={
		\draw[mark repeat=2,mark phase=2]
		plot coordinates {
			(0cm,0cm)
			(0.3cm,0cm)        
			(0.6cm,0cm)         
		};%
	}
}

\pgfplotstableread[col sep = tab]
    {Data/inference_gpu.log}
    \tableGPU

\pgfplotstableread[col sep = tab]
    {Data/inference_asc-1c-2t.log}
    \tableTPU

\pgfplotstableread[col sep = tab]
    {Data/inference_asc-4c-2t.log}
    \tableTPUtwo

\pgfplotstableread[col sep = tab]
    {Data/inference_cpu-1c.log}
    \tableCPU

\pgfplotstableread[col sep = tab]
    {Data/inference_cpu-52c.log}
    \tableCPUtwo
\centering
\subcaptionbox{Batch processing and completion delay vs. batch size.\label{fig:hw_delay}}
	[0.32\textwidth] 
	{\begin{tikzpicture}
    \begin{axis}[
        xmode=log, log ticks with fixed point,
        scaled x ticks=false,
        ylabel near ticks,
        ymin=0, ymax=100,
        xmin=8, xmax=16384,
    xtick=data,
    /pgf/number format/.cd,fixed,precision=0,
    xticklabel style={
        rotate=45,
        scaled ticks=false
        },
    xticklabel=\pgfmathparse{exp(\tick)}\pgfmathprintnumber{\pgfmathresult},
        ylabel={Delay [ms]},
        scale only axis,
        width=\xrat*\textwidth, 
        height =\yrat*\heightfigure,
        xlabel = Batch size,
        ylabel style={align=center},
        yticklabel style={
            font = \footnotesize
        },
        xticklabel style={
            font = \footnotesize
        },
        label style = {
            font=\footnotesize
        },
        ylabel style = {
            font = \footnotesize,
            inner sep = 1pt,
            at={(axis description cs:-0.1,0.45)}
        },
        x label style={
            font=\footnotesize,
            at={(axis description cs:0.5,\xlab)},
            anchor=north},
        legend style={
            font=\scriptsize,
            at={(1.0, 1.01)},
            anchor = south east,
            draw = none,
            fill = white,
            inner sep=1pt,
            /tikz/every even column/.append style={column sep=3pt}
        },
        legend pos = north west,
        legend columns=1,
        legend cell align={left},
        ymajorgrids,
        xmajorgrids,
        yminorgrids,
        enlargelimits=0.02,
        cycle list name=BenchmarkCycleList
    ]

    \addplot+ [] table [
                x index = 0, 
                y expr = ((1000*\thisrowno{0} / \thisrowno{1})) 
            ] {\tableGPU};

    \addplot+ [] table [
                x index = 0, 
                y expr = ((1000*\thisrowno{0} / \thisrowno{1})) 
            ] {\tableTPU}; 
            
    \addplot+ [] table [
                x index = 0, 
                y expr = ((1000*\thisrowno{0} / \thisrowno{1})) 
            ] {\tableTPUtwo};
            
    \addplot+ [] table [
                x index = 0, 
                y expr = ((1000*\thisrowno{0} / \thisrowno{1})) 
            ] {\tableCPU};
            
    \addplot+ [] table [
                x index = 0, 
                y expr = ((1000*\thisrowno{0} /\thisrowno{1})) 
            ] {\tableCPUtwo};
    
    \addplot+ [] table [
                x index = 0, 
                y expr = ((\thisrowno{0} /\relativeratek)) 
            ] {\tableCPUtwo};
    \legend{GPU, TPU-1, TPU-4, CPU-1, CPU-52}
    \node[draw=black,
        fill=white,
        inner sep=2pt,
        anchor=south east,
        align=center,
        text width = 4em,
        rounded corners,
        thin,
        font=\tiny,
        ]
        at (rel axis cs: 0.7, 0.5)
        (compltetion)
        {Batch completion delay (\SI{\relativeratek}{\kilo class/s})};
\draw [-latex]
    (compltetion) to
    (axis cs: 1900, 42);
    \end{axis}
\end{tikzpicture}}
\hfill%
\subcaptionbox{Ratio of dead short-lived flows after the \ith{10} packet.\label{fig:dead}}
	[0.28\textwidth]
	{\begin{tikzpicture}
    \begin{axis}[
        scale only axis,
        ymin=0, ymax=1,
        xmin=0.1, xmax=70,
        width = 0.22*\textwidth, 
        height = \yrat*\heightfigure,
        xlabel = {Time since \ith{10} packet [ms]},
        ylabel = Ratio of dead flows,
        label style = {
            font=\footnotesize
        },
        legend style={
            font=\tiny,
            draw = none,
            inner sep = 0pt
        },
        x tick label style={
            font = \footnotesize
        },
        legend pos = north west,
        legend columns=1,
        legend cell align={left},
        ymajorgrids,
        xmajorgrids,
        enlargelimits=0.02,
        cycle list name=MyCyclelist
    ]
    \pgfplotstableread[col sep = comma]
    {Data/dead_short_flows.csv}
    \mytable
    
        \addplot+ [] table [
                x = delays, 
                y = dead_transit 
            ] {\mytable};
            \addlegendentry{Transit}
        \addplot+ [] table [
                x = delays, 
                y = dead_access
            ] {\mytable};
            \addlegendentry{Access}
        \addplot+ [] table [
                x = delays, 
                y = dead_campus 
            ] {\mytable};
            \addlegendentry{Campus}


    \end{axis}
\end{tikzpicture}}
\hfill%
\subcaptionbox{Number of received packets for long-lived flows after the \ith{10} packet.\label{fig:shift}}
	[0.36\textwidth]
	{\def\rat{0.13}
\begin{tikzpicture}
    \begin{axis}[
        scale only axis,
        boxplot/draw direction=y,
        boxplot = {
            draw position={1/4 + floor(\plotnumofactualtype/3) + 1/4*mod(\plotnumofactualtype,3)},
            box extend=0.19,
        },
        ymin=1, ymax=105,
        xmin=0, xmax=5,
        width= \xrat*\textwidth, 
        height = \yrat*\heightfigure,
        xlabel = {Time since \ith{10} packet [ms]},
        ylabel = Nb of packets per flow,
        ylabel style = {
            align = center,
            font=\footnotesize,
            at={(axis description cs:-0.1,.5)}},
        xlabel style = {
            font=\footnotesize},
        ytick={0, 25, ..., 100},
        x=\rat*\columnwidth, 
        xticklabels={
            {0},
            {1},%
            {10},%
            {20},%
            {30},%
            {40}},
        x tick label style={
            text width=0.05*\columnwidth,
            align=center,
            font = \footnotesize
        },
        xtick distance=1,
        x tick label as interval,
        legend style={
            font = \tiny,
            draw = none,
            fill = none,
            inner sep = 1pt,
            at = {(0.5,1.02)},
            anchor = south
        },
        legend pos = north west, 
        legend columns=1,
        legend cell align={left},
        boxplot legend,
        ymajorgrids,
        enlarge y limits=0.01,
        enlarge x limits=0,
        cycle list name=uniformCyclelistFilled
    ]
    \pgfplotstableread[col sep = comma]
    {Data/untagged_long_flows.csv}
    \mytable

    \pgfplotstablegetrowsof{\mytable}
    \pgfmathsetmacro\numberofrows{\pgfplotsretval-1}

    \pgfplotsinvokeforeach{0,...,\numberofrows}{
        \addplot+[
            boxplot prepared from table={
                table=\mytable,
                row =#1,
                lower whisker=min_transit,
                upper whisker=max_transit,
                lower quartile=lw_transit,
                upper quartile=uw_transit,
                median=med_transit
                }, 
            boxplot prepared]
        coordinates {};
        \addplot+[
            boxplot prepared from table={
                table=\mytable,
                row =#1,
                lower whisker=min_access,
                upper whisker=max_access,
                lower quartile=lw_access,
                upper quartile=uw_access,
                median=med_access
                }, 
            boxplot prepared]
        coordinates {};
        \addplot+[
            boxplot prepared from table={
                table=\mytable,
                row =#1,
                lower whisker=min_campus,
                upper whisker=max_campus,
                lower quartile=lw_campus,
                upper quartile=uw_campus,
                median=med_campus
                }, 
            boxplot prepared]
        coordinates {};
    }
    \legend{Transit, Access, Campus}
    \end{axis}
\end{tikzpicture}}
\hfill%

\caption{Delay requirements. Boxplots show \ith{99}, \ith{75}, \ith{25}, and $1^{\text{st}}$ percentiles.}
\end{figure*}
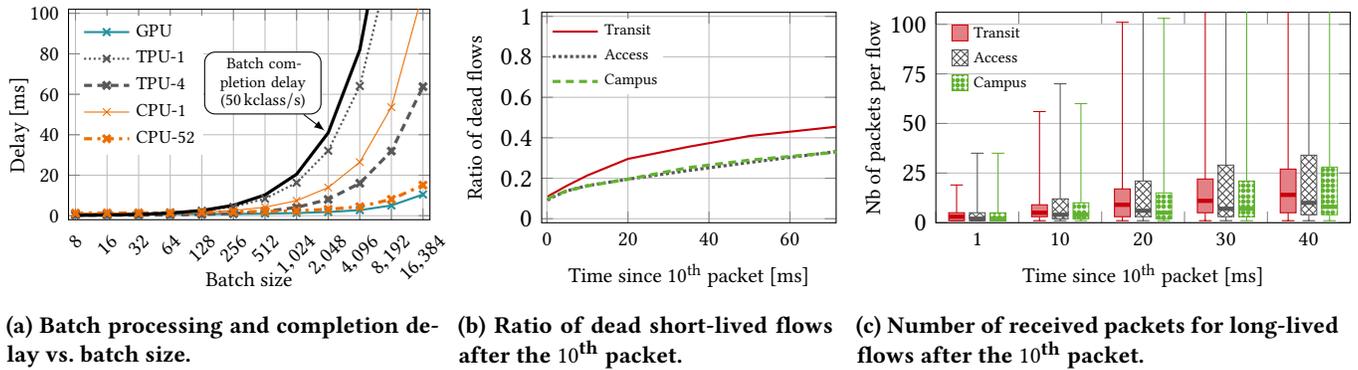

\subsection{Throughput}
The capacity of a network device to sustain a given throughput is an essential feature, not only from the traditional perspective of data rate (measured in bit per second) but also from the perspective of flow analytics (measured in classification per second). \acrshort{fix} design should aim to not sacrifice the former to perform the latter.
We also have to consider the foreseen growth of both data rates (due to higher throughput in local networks) and flow analytics (due to the increase of connected devices). While today's edge routers typically deal with throughput in the order of single-digit Gigabits per second (Gbps), the requirements for the next-generation routers consider up to \SI{100}{Gbps}. In the following, to put \acrshort{fix} under stress, we will use this configuration for \SI{100}{Gbps} network routers as a reference.

We show in \Cref{tab:dataset} the total number of packets, flows and series (\ie{} flows with more than ten packets) in our datasets. We use these traffic characteristics to derive the required performance in terms of packet (\ie{} Mpps) and series processing speed (\ie{} class./s) for the reference \SI{100}{\giga{}bps} at maximum load. We conclude that \acrshort{fix} deployed in a single \SI{100}{\giga{}bps} linecard should be able to sustain \SI{15}{Mpps} and \SI{50}{kclass/s}. 

We then evaluate whether the hardware accelerator can sustain such an analytics rate. We report analytics throughput in \Cref{fig:hw_classrate} in terms of classification per second achieved by CPU (both 1 and 52 cores), GPU (640 cores available), and TPU (both 1 and 4 cores). The main parameter that impacts the analytics rate is the \emph{batch size}, \ie{} the number of series that are grouped for parallel processing. To evaluate this aspect, we performed a \SI{60}{s} long stress test by submitting a stream of inference tasks of a given batch size, 
while measuring the time between each submission and the related output delivery. We show in \Cref{fig:hw_classrate} that, with a batch size greater than \num{64}, every hardware accelerator can sustain a target \SI{50}{\kilo class/s}. Interestingly, whereas CPU and GPU are optimized for large batch size, the 4 TPU chips combined are significantly faster in processing small batches than both 52 CPUs and GPU.

The stress test however does not correctly represent real traffic behavior. Network traffic is subject to \emph{instantaneous} high-load over a short period, \ie{} bursts~\cite{JinGXSNLQW19}, and this can be a challenge regarding the overall system throughput. To estimate whether incoming data bursts also reflect in the series arrival rate, we measured during \SI{10}{\minute} of each dataset the \textit{batch completion delay}, which is the time needed to complete a batch of a given size. Thus, for a batch $B=8$ it is the time needed for \num{8} consecutive new series to enter in the system.
We computed the ratio between the longest and the shortest completion delays observed within each second.  
A high ratio indicates that the batch completion delays are subject to a high variation. Results in \Cref{fig:bursts} show that batch completion delays vary significantly, especially with small batches. Hence, we highlight that the results in \Cref{fig:hw_classrate} on the maximum classification rate for a given configuration addresses one aspect of the analytics rate while \acrshort{fix} should also be able to absorb bursts, by dynamically adapting the batch sizes to sudden series arrivals.

Finally, we analyze the throughput with respect to energy consumption, which is a key performance criterion in today's edge devices. \Cref{fig:hw_quadrant} illustrates the tradeoff of analytics throughput versus power usage for varying batch sizes as a scatter plot of the power ($x$-axis) and classification rate ($y$-axis) normalized with respect to reference values of \SI{\relativepower}{\watt} that is the power consumption of an idle \gls{gpu} and \SI{\relativeratek}{\kilo class/s}. The top-left square highlights desirable configurations; the bottom-right square includes operational regions that should be avoided. Notice that only when using a single TPU we meet our requirements (\SI{\relativepower}{\watt} and \SI{\relativeratek}{\kilo class/s}) when using a couple of configurations, while all other scenarios fail in at least one dimension.

\input{Figures/modular.tex}

\subsection{Delay}
The second essential feature of a traffic analytics pipeline is the delay between the time at which the analytics can be processed (in our case study, it is the time at which the \ith{K} packet arrives at the device) and the time at which the result of the analytics can be exploited. 

More in details, we split delay into two components: the \textit{analytics processing delay} and the \textit{batch completion delay}. We report in \Cref{fig:hw_delay} the analytics processing delay for multiple batch sizes. We emphasize here the trade-off between throughput and delay since the results from \Cref{fig:hw_classrate} call for the use of a larger batch, but the processing of a large batch generates a significant delay. Furthermore, we added in \Cref{fig:hw_delay} a line to represent the batch completion delay for a \SI{50}{\kilo class/s} constant series arrival rate. The batch completion delay is dominant in comparison to the analytics processing delay. We conclude that setting the batch size is key: a large batch results in a long delay, while a small one leads the inference system to run in a sub-optimal operational regime.

We highlight now the impact of delay in flow analytics when regarded from the application of the classification result. We distinguish flows in \emph{short-lived} and \emph{long-lived}. The former represents the vast majority of the flows while the latter is responsible for the vast majority of data volume. In our datasets, considering the target of analytics after the \ith{10} packet we identify a flow as short-lived if it has less than 35 packets, long-lived otherwise. \acrshort{fix} targets to identify the application label for a short-lived flow before observing the flow last packet to avoid \emph{post-mortem} analytics. Indeed, such a label is meant to tag flow's packets before forwarding them. However, the life expectancy of short-lived flows after the \ith{10} packet arrival ranges from a few milliseconds to minutes (see \Cref{fig:dead}). Long-lived flows on the contrary are likely classified before their end, but \acrshort{fix} targets early classification to enable effective traffic differentiation~\cite{LiNCGM19}, so the sooner the classification, the better.

\Cref{fig:shift} presents the number of packets received by long-lived flows since the \ith{10} packet. Waiting for the label drives the system to an increasing number of untagged packets hence deteriorating the performance of the algorithm exploiting such information. 
We conclude that delay requirement cannot be strictly and universally set, since it also depends on the location of the device in the network (see the differences between datasets). Yet, from our analysis, a target around \SI{10}{\milli\second}  enables a low fraction of both untagged packets and post-mortem analytics.

\subsection{Takeaway}
We showed in this section that the design of \acrshort{fix} should take into account multiple, sometimes conflicting, objectives related to throughput, delay, and energy consumption. 

We are positive in the capabilities of hardware accelerators to handle the classification rates for regular edge routers, even at \SI{100}{Gbps}, but we noticed that tailored policies regarding batch size have to be designed to deal with bursts. Whereas it can meet the throughput requirements, the CPU does not offer a viable hardware platform for fast analytics since the energy consumption is too large while it does not offer processing gains. TPU offers a more energy-efficient alternative, especially with regards to its higher processing performance for small batches, while GPU is the option to adopt for large throughput requirements.

The study of \gls{iad} in flows revealed that we cannot express any unique delay objective, contrarily to previous work on inference service for image processing applications~\cite{clipper}. We recommend small batch sizes to meet a delay objective around \SI{10}{\milli\second}, but we also showed that small batch sizes cannot sustain high throughput nor low power consumption. These three objectives (throughput, delay, and energy) define the operational points of the \acrshort{fix} system.

\section{FENXI Overview}

This section introduces \acrshort{fix} design, which is presented in \Cref{fig:modular}. The \acrshort{fix} architecture splits the processing between \textit{Flow managers} and \textit{Analytics managers}, and each manager can be deployed on dedicated independent processing units. This design choice is driven by the fact that the two managers operate at different granularities. 
Flow managers handle packet entering the system, group them according to the related flow, and extract features related to the first $K$ packets of each flow; Analytics managers receive per-flow features from Flow managers, further group them into batches, and trigger DL inference on a batch.

A Flow and an Analytics manager together constitute a single processing pipeline and communicate through a communication ring \textit{cRing}, which enables zero-copy and lock-less data exchange. In the following, we introduce \acrshort{fix} building blocks that we will detail in the rest of the paper.




\paragraph{Flow Manager} The flow manager is \acrshort{fix}
interface towards the system NICs. Received packets are processed according to the steps pictured on the left side of \Cref{fig:modular}. 
\acrshort{fix} identifies the flow through the IP \num{5}-tuple. If the flow has been already classified in the past, it forwards the packet with a tag indicating the classification result. If the flow has not been classified yet, \acrshort{fix} updates its internal data structures and forwards the packet with an empty tag. Furthermore, if the received packet completes a series \ie{} this packet is the \ith{K} packet of this flow, \acrshort{fix} forwards the extracted features and the flow tuple to the next element in the pipeline (the analytics manager) via the cRing.

To fulfill device requirements (cf. \Cref{sec:rqrmnt}%
), the flow manager must reduce per-packet processing overhead, such that any function carefully considers hardware and software capabilities. As stated in \Cref{sec:related}, previous work on \textit{offloaded} inferring services have either not addressed line rate pre-processing (for example in the literature on image processing~\cite{ZhangYWY19,LeeSCSWI18}) or considered less demanding offloading tasks (for example load-balancing~\cite{JinGXSNLQW19}). Feature extraction at line rate is critical for in-device analytics implementation. We describe the feature extraction process in \Cref{sec:feature}.

\paragraph{Analytics Manager}
The Analytics manager constantly polls the cRing, waiting for flow tuple and series to be pulled. To increase efficiency, the software operations of the Analytics manager are tightly coupled to the underlying \gls{dl} hardware accelerator. To fulfill constraints and requirements expressed in 
\Cref{sec:rqrmnt}, %
we further split the Analytics manager into two sub-modules \textit{Batching} and \textit{Caching} with the twofold objective of improving data transmission and analytics processing speed. Batching is responsible for building groups of series for which analytics will be computed in parallel. Caching acts as a filter before batch composition and stores recently executed analytics to speed up processing. 

Each series is processed by the Analytics manager according to the steps pictured on the right side of \Cref{fig:modular}. If the series is cached, the result is immediately returned to the Flow manager. Otherwise, a batch of series is formed and, when ready, sent to the Analytics device for processing. In turn, when the result of the batch is ready, the analytics manager forwards the labels back to the Flow manager.

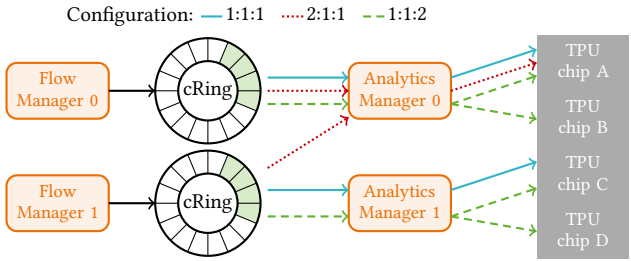
\begin{figure}[t!]
    \centering
    \pgfdeclarelayer{background}
\pgfdeclarelayer{foreground}
\pgfsetlayers{background,main,foreground}

\begin{tikzpicture}

\tikzset{%
    legarrow/.style={
        font=\footnotesize,
        inner sep=1pt,
        fill = white,
    },
    arr/.style={
        ->, thick
    }, 
    oneoneone/.style={
        colorBlue
    }, 
    twooneone/.style={
        colorRed, densely dotted
    }, 
    oneonetwo/.style={
        colorGreen, densely dashed
    }, 
    pics/manager/.style n args={1}{ 
        code = {
            \node[rectangle,
                rounded corners,
                text width = 0.8*\sizecomponenttext,
                minimum height = \heightcomponent,
                font = \scriptsize,
                rounded corners,
                thick,
                align = center,
                fill=colorOrange!10,
                draw=colorOrange,
                anchor=center,
                ]
                {\color{colorOrange}#1};
            }
    }, 
    pics/packet/.style n args={3}{ 
        code = {
            \node[rectangle,
                rounded corners,
                text width = 0.4*\sizecomponenttext,
                minimum height = \heightcomponent,
                fill=#2,
                draw=#3,
                very thick,
                anchor=center,
                ]
                (letter)
                { };
            \draw[
                color=#3,
                thick]
                ([yshift=-1.5pt, xshift=-1.5pt]letter.35) -- (0,0);
            \draw[
                color=#3,
                thick]
                ([yshift=-1.5pt, xshift=1.5pt]letter.145) -- (0,0);
            \node[
                text width = \sizecomponenttext,
                minimum height = \heightcomponent,
                draw=none,
                anchor=center,
                font=\scriptsize,
                align = center]
                (letter)
                {\textbf{#1}};
        }
    },
    chip/.style={
                text width = 0.7*\sizecomponenttext,
                minimum height = 0.8*\heightcomponent,
                draw=none,
                font=\scriptsize,
                fill=colorGray!50,
                align = center
    }, 
    pics/circular/.style n args={1}{
        code = {
            \fill [colorGreen!25] 
                (0,0) -- 
                (67.5:#1) arc [end angle=-22.5, start angle=67.5, radius=#1] -- 
                cycle;
            \draw [thick] (0,0) circle (#1);
            \foreach \angle in {90,67.5,...,-67.5}
                \draw (\angle:#1) -- (\angle-180:#1);
            \node [circle,
                    font = \footnotesize,
                    thick,
                    fill=white,
                    draw=black,
                    align=center,
                    inner sep = 1pt,
                    anchor = center] 
                at (0,0) 
                {cRing};
        }
    }
}


\def\sizecomponenttext{4em}
\def\heightcomponent{2em}

\def\shiftx{5em}
\def\shifty{3.2em}
\def\shi{15pt}


\pic[local bounding box = flowzero]
    {manager={Flow Manager 0}{}};

\pic[local bounding box = cringzero,
    right=0.75*\shiftx of flowzero]
        {circular = {0.7}{}};

\pic[local bounding box = analyticszero,
    right=\shiftx of cringzero]
    {manager={Analytics Manager 0}{}};
    
\pic[local bounding box = flowone,
    below = \shifty of flowzero]
    {manager={Flow Manager 1}{}};

\pic[local bounding box = cringone,
    right=0.75*\shiftx of flowone]
        {circular = {0.7}{}};
        
\pic[local bounding box = analyticsone,
    right=\shiftx of cringone]
    {manager={Analytics Manager 1}{}};
    
\node[chip, anchor=south]
    at ([xshift=\shiftx]
        analyticszero.east)
    (chipA)
    {\color{white}TPU chip A};

\node[chip, anchor=north]
    at ([xshift=\shiftx]
        analyticszero.east)
    (chipB)
    {\color{white}TPU chip B};

\node[chip, anchor=south]
    at ([xshift=\shiftx]
        analyticsone.east)
    (chipC)
    {\color{white}TPU chip C};

\node[chip, anchor=north]
    at ([xshift=\shiftx]
        analyticsone.east)
    (chipD)
    {\color{white}TPU chip D};


\draw[arr, black]
    (flowzero) to
    (cringzero);
    
\draw[arr, black]
    (flowone) to
    (cringone);

\def\yshi{5pt}
\draw[arr, oneoneone]
    ([yshift=\yshi] cringzero.east) to
    ([yshift=\yshi] analyticszero.west);

\draw[arr, twooneone]
    (cringzero) to
    (analyticszero.west);

\draw[arr, oneonetwo]
    ([yshift=-\yshi] cringzero.east) to
    ([yshift=-\yshi] analyticszero.west);

\draw[arr, twooneone]
    (cringone) to
    ([yshift=-2*\yshi] analyticszero.west);
    
\draw[arr, oneoneone]
    ([yshift=\yshi] cringone.east) to
    ([yshift=\yshi] analyticsone.west);
    
\draw[arr, oneonetwo]
    ([yshift=-\yshi] cringone.east) to
    ([yshift=-\yshi] analyticsone.west);


\draw[arr, oneoneone]
    ([yshift=\yshi] analyticszero.east) to
    ([yshift=\yshi] chipA.west);

\draw[arr, twooneone]
    (analyticszero.east) to
    (chipA.west);

\draw[arr, oneonetwo]
    ([yshift=-\yshi] analyticszero.east) to
    ([yshift=-\yshi] chipA.west);

\draw[arr, oneonetwo]
    ([yshift=-\yshi] analyticszero.east) to
    (chipB.west);
    
\draw[arr, oneoneone]
    ([yshift=\yshi] analyticsone.east) to
    ([yshift=\yshi] chipC.west);
    
\draw[arr, oneonetwo]
    ([yshift=-\yshi] analyticsone.east) to
    ([yshift=-\yshi] chipC.west);

\draw[arr, oneonetwo]
    ([yshift=-\yshi] analyticsone.east) to
    (chipD.west);
    
\def\sizeleg{8pt}
\def\colsep{2.2em}

\node[anchor=south west, 
        font=\footnotesize]
    at ([yshift=1pt]
        flowzero |- cringzero.north)
    (conf)
    {Configuration:};

\node[legarrow]
    at ([xshift=2*\sizeleg]conf.east)
    (one)
    {1:1:1\vphantom{g}};
\draw[oneoneone, thick]
    (one.west) to
    ([xshift = -\sizeleg] one.west);
    
\node[legarrow]
    at ([xshift=\colsep] one.east)
    (two)
    {2:1:1\vphantom{g}};
\draw[twooneone, thick]
    (two.west) to
    ([xshift = -\sizeleg] two.west);
    
\node[legarrow]
    at ([xshift=\colsep] two.east)
    (three)
    {1:1:2\vphantom{g}};
\draw[oneonetwo, thick]
    (three.west) to
    ([xshift = -\sizeleg] three.west);

\end{tikzpicture}
    \caption{Pipeline deployment strategies.}
    \label{fig:dual_pipeline}
\end{figure}

\paragraph{Pipeline deployment strategies}
We showed in \Cref{sec:rqrmnt} that different networks present different packets, flows, and series rates. Hence, the two pipeline stages can be more or less latency and computational processing-demanding, and bottlenecks do not necessarily happen at the same stage of the pipeline. As such, we design \acrshort{fix} to have a configurable multi-pipeline system. 
To reduce system overhead, we only consider lock-free strategies for which the same flow/series information is always handled by the same couple of Flow and Analytics managers, which run on separate processing units. 

Thanks to its flexible design \acrshort{fix} pipeline can be configured to address the processing bottleneck by means of deployment strategies. \Cref{fig:dual_pipeline} presents three lock-free multi-pipeline deployment strategies. $(i)$ A pipeline in a 1:1:1 configuration is the default configuration with a Flow manager, an Analytics manager, and an analytics processor \ie{} the \gls{tpu} that runs in isolation. $(ii)$ The 1:1:2 configuration addresses a bottleneck in the analytics processing throughput. In this configuration, a single Analytics manager balances the analytics load across more than one \gls{tpu} chips. $(iii)$ The 2:1:1 configuration copes with scenarios with high packet arrival rates but mid/low series arrival rates. In such a case a single Analytics manager interacting with a single \gls{tpu} can retrieve series from multiple Flow managers. 

Finally, for each strategy, \acrshort{fix} can scale up by instantiating multiple pipelines in parallel by exploiting modern NICs dynamic forwarding of incoming traffic into different receive queues by using \gls{rss} to load balance packets acquisition across different Flow manages. 

According to our analysis in \Cref{sec:rqrmnt}, a single \gls{tpu} can sustain the load for \SI{100}{\giga{} bps}. Nonetheless, we deploy a system with the multi-pipeline 1:1:1 configuration, which maintains a lower load on the hardware accelerator and can absorb instantaneous high series arrival rate over short periods.

\section{Flow Manager}\label{sec:feature}
The Flow manager module aims to extract the features that are relevant for the analytics. A high-level diagram of its operations is presented in \Cref{fig:modular}. Packets are split into flows according to the classic IP \num{5}-tuple and then processed depending on three cases. $(i)$ The packet is part of a \emph{labeled} flow, \ie{} \acrshort{fix} previously ran analytics on this flow, which resulted in a label. In this case, the forwarding plane performs actions on the packet with respect to the label, \eg{} \emph{tagging} the packet before forwarding it. It is important to highlight that when the packet is part of an unlabeled flow, rather than waiting for the label to be computed, \acrshort{fix} forwards the packet unmodified. We say that the packet is \textit{untagged}. In the latter scenario, we further distinguish two cases: $(ii)$ if the packet is not the \ith{K} packet of the flow, \acrshort{fix} only updates the flow state; $(iii)$ If the packet is the \ith{K} packet of the flow, the analytics processing for this flow can be triggered. \acrshort{fix} forwards the \emph{features} extracted from the first $K$ packets (\eg{} a series of packet properties), and passes it to the next pipeline element, which asynchronously retrieves it via the cRing. 

\subsection{System Design}
The Flow manager data plane is designed around \gls{dpdk} but, in principle, it can be ported to similar packet processing frameworks for general-purpose servers~\cite{netmap} or smart NICs~\cite{mellanoxnic, huaweinic}.

\begin{figure}[t!]
    \centering
    \input{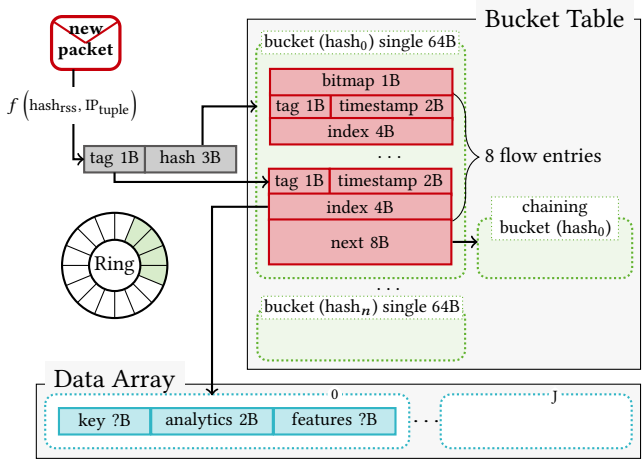}
    \caption{Flow manager Hash table.}
    \label{fig:hash_table}
\end{figure}

At startup time, the Flow manager instantiates several workers, one for each pipeline, which independently handles ongoing flows dispatched by the linecard with \gls{rss}. To keep track of ongoing flows, \acrshort{fix} uses a hash table (see \Cref{fig:hash_table}) with multiple entries buckets (\ie{} 8) coupled with a data array, which is addressed using array indexes retrieved through a dedicated ring buffer. The data array size (\ie{} the maximum number of entries that can be stored by the flow manager) and the number of buckets can differ and hence used to artificially control the number of collisions within a single bucket at the expense of memory efficiency.

When a packet arrives, the hash table bucket position is computed using the three least significant bytes of a \textit{hash} value computed on the IP \num{5}-tuple. To reduce as much as possible the packet processing time the Flow manager relies on the fact that modern linecards and NICs compute symmetric hash values \cite{symmtoeplitz} (\ie{} the same hash is computed for both directions of a flow) on the IP \num{5}-tuple to load balance the flows across the different receive queues \ie{} \gls{rss}, and attach such value to packet's metadata. To mitigate the unbalance issues reported by \citet{symmtoeplitz} when using the Toeplitz symmetric hash function we compute our hash as: 
\[f(hash_{rss},IP_{tuple})=hash_{rss} \oplus IP_{src} \oplus IP_{dst}\]
where $\oplus$ represents the bitwise XOR operation, which provides a balanced hash function at a very low processing cost.

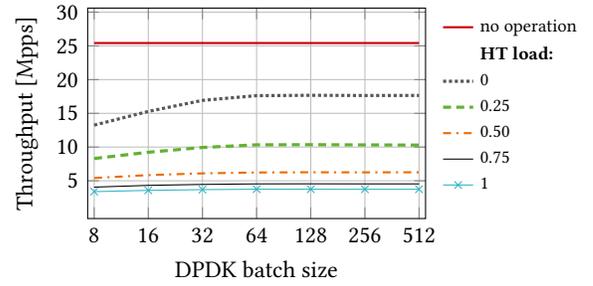
\begin{figure}[t!]
    \centering
    \pgfplotsset{
	compat=1.11,
	legend image code/.code={
		\draw[mark repeat=2,mark phase=2]
		plot coordinates {
			(0cm,0cm)
			(0.2cm,0cm)        
			(0.4cm,0cm)         
		};%
	}
}

\begin{tikzpicture}
    \begin{axis}[
        xmode=log, log ticks with fixed point,
        scale only axis,
        ymin=0, ymax=30,
        xtick=data,
        ytick={5,10,15,20,25,30},
        width= 0.53\columnwidth, 
        height = \heightfigure,
        xlabel = DPDK batch size,
        ylabel = {Throughput [Mpps]},
        label style = {
            font=\small
        },
        legend style={
            font=\scriptsize,
            draw = none,
            inner sep = 2pt,
        },
        x tick label style={
            font = \small
        },
        y tick label style={
            font = \small
        },        
        legend pos = outer north east,
        legend columns=1,
        legend cell align={left},
        ymajorgrids,
        xmajorgrids,
        enlargelimits=0.02,
        cycle list name=MyCyclelist
    ]
    \pgfplotstableread[col sep=comma]
    {Data/FM_cores_batch.csv}
    \mytable


    \addplot+ []coordinates{(8,25.42)(16,25.42)(32,25.42)(64,25.42)(128,25.42)(256,25.42)(512,25.42)};
    \addlegendimage{empty legend}

    \pgfplotsinvokeforeach{
        0,
        0.25,
        0.5,
        0.75,
        1}{
        
        \addplot+ [] table [
                x = batchsize, 
                y = load_#1
            ] {\mytable};
            
        }
    
        \legend{
            no operation,
            \textbf{HT load:},
            0,
            0.25,
            0.50,
            0.75,
            1}
    
    \end{axis}
\end{tikzpicture}
    \caption{Single worker flow manager processing speed with increasing batch size and hash table load.}
    \label{fig:dpdk-batch}
\end{figure}

The hash table bucket is stored into a single cache line (\ie{} \SI{64}{Bytes}) and composed by a \textit{bitmap} used to quickly check the occupied entries in the bucket, eight flow entries, and a \textit{next} value, which points to an external, dynamically allocated, memory area to handle chaining. Each flow entry stores a \SI{1}{Byte} \textit{tag} extracted from the hash value, a \SI{2}{Bytes} \textit{timestamp}, and a \SI{4}{Bytes} \textit{index}. We use the hash tag, which corresponds to the first byte of the hash value itself, for a quick comparison to accelerate lookups in case of a high number of collisions~\cite{memc3}. The coarse grain timestamp (in seconds) stores the last time the flow entry was accessed (\eg{} for packet labeling or statistics update) to lazily remove inactive flow entries based on the \textit{stale} timeout system parameter. Finally, we use the \textit{index} to point to the element of the data array that stores the information of the flow. A separate ring buffer is used to keep track of data array indexes, which are available for storing flow information.
                                
The data array stores $(i)$ the flow's \textit{key} (\ie{} the IP \num{5}-tuple used during the lookup), $(ii)$ the result from the analytics  (\ie{}  the \textit{label} stored as an atomic variable to avoid contention), $(iii)$ the features extracted from the flow (\ie{} packet size and direction of the first $K$ packets), and $(iv)$ some statistics about the flow (\eg{} the number of packets).

\subsection{Micro-benchmark}
We ran two micro-benchmarks to better understand \acrshort{fix} Flow manager performance in isolation (without the Analytics manager), and in worst-case conditions, \SI{64}{Bytes} packets, using two servers with DPDK 20.02 (see the technical description in \Cref{sec:rqrmnt}), which are directly connected via a \SI{100}{Gbps} link. 

In the first test, we evaluated the processing speed of the Flow manager with a single worker for increasing DPDK batch size \ie{} the maximum number of packets the system will receive in parallel. The data array size is fixed to $2^{19}$ (\SI{524}{k}) with $2^{17}$ (\SI{131}{\kilo\nothing}) buckets. At startup, we pre-loaded the hash table to reach a fixed load factor ($load=elements/size(data\_ array)$). During a \SI{30}{\second} time frame, we sent \SI{64}{Bytes} packets belonging to the pre-loaded flows at maximum speed to the Flow manager from the second server running Moongen~\cite{moongen}. \Cref{fig:dpdk-batch} shows the maximum number of packets retrieved in parallel by the DPDK framework for different hash table loads (at load 0,  a single flow is used during the test). The \textit{no operation} reference corresponds to Flow manager workers that receive and forward packets without performing any additional operation. Large batch sizes improve the packet processing efficiency, leading to a higher speed, especially for lower loads, which are not far from the limit highlighted by the \textit{no operation} reference 
line. Moreover, higher hash table loads lead to slower processing speed, which highlights the fact that a hash table set with a wrong dimension can become a bottleneck. 

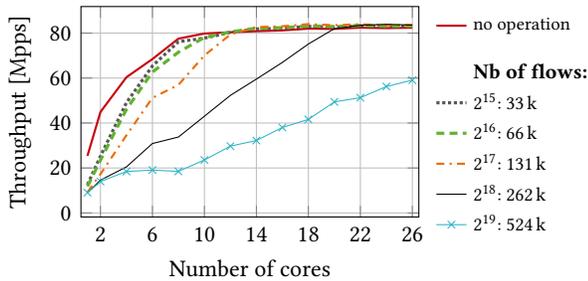
\begin{figure}[t!]
    \centering
    \pgfplotsset{
	compat=1.11,
	legend image code/.code={
		\draw[mark repeat=2,mark phase=2]
		plot coordinates {
			(0cm,0cm)
			(0.2cm,0cm)        
			(0.4cm,0cm)         
		};%
	}
}

\begin{tikzpicture}
    \begin{axis}[
        scale only axis,
        ymin=0, ymax=90,
        xmin=1, xmax=26,
        width= 0.53\columnwidth, 
        height = \heightfigure,
        xlabel = Number of cores,
        ylabel = {Throughput [Mpps]},
        xtick={2, 6, ..., 26},
        label style = {
            font=\small
        },
        legend style={
            font=\scriptsize,
            draw = none,
            inner sep = 2pt,
            /tikz/row 2/.style={
                column sep=-5pt, 
                text height=3ex,
                font=\footnotesize},
        },
        x tick label style={
            font = \small
        },
        y tick label style={
            font = \small
        },        
        legend pos = outer north east,
        legend columns=1,
        legend cell align={left},
        ymajorgrids,
        xmajorgrids,
        enlargelimits=0.02,
        cycle list name=MyCyclelist
    ]
    \pgfplotstableread[col sep=comma]
    {Data/FM_load_cores.csv}
    \mytable

    \addplot+ [] table [
                x = cores, 
                y = load_0
            ] {\mytable};
    
    \addlegendimage{empty legend}

    \pgfplotsinvokeforeach{
        16384,
        32768, 
        65536,
        131072, 
        262144}{
        \addplot+ [] table [
                x = cores, 
                y = load_#1
            ] {\mytable};
        }

    \legend{
        no operation,
        \textbf{Nb of flows:},
        $2^{15}$: \SI{33}{\kilo\nothing},
        $2^{16}$: \SI{66}{\kilo\nothing},
        $2^{17}$: \SI{131}{\kilo\nothing},
        $2^{18}$: \SI{262}{\kilo\nothing},
        $2^{19}$:
        \SI{524}{\kilo\nothing}}
    
    \end{axis}
\end{tikzpicture}
    \caption{Flow manager scalability at  hash table load 0.5 with increasing number of workers and flows.}
    \label{fig:dpdk-cores}
\end{figure}

In the second test, we evaluated the processing speed with an increasing number of Flow manager workers (one per core) and a different number of flows. We pre-loaded the hash table such that it reached a \num{0.5} load factor and sent \SI{64}{Bytes} packets belonging to such flows. We present in \Cref{fig:dpdk-cores} the processing speed for a bucket size that corresponds to a quarter of the size of the hash table. The more workers are used by the flow manager, the higher is the processing speed. Moreover, processing speed is limited at \SI{80}{ Mpps} due to hardware limitations (the theoretical limit is \SI{144.88}{Mpps}) imposed by the PCI Express 3.0 x16, which our \SI{100}{Gbps} NIC is attached to. Similar limitations are reported by \citet{pcie}.
We also observe that the processing speed does not only depend on the hash table load factor but also on the number of concurrent flows. Hence, the hash table timeout for removing old elements needs to be correctly set to avoid too many stale flows, \eg{} \SI{30}{\second}.

From our micro-benchmarks and the preliminary system requirements analysis in 
\Cref{sec:rqrmnt}
we conclude that to sustain \SI{100}{Gbps} (about \SI{13}{Mpps} in the considered datasets) two-three cores suffice with a properly sized flow table.

\section{Analytics Manager}\label{sec:analytics}
The analytics manager module takes as input a tuple (IP 5-tuple, packet time series) generated by a Flow manager and perform the desired analytics \ie{} early flow classification. Similarly to the flow manager, at startup, the analytics manager instantiates several workers, one per pipeline according to the multi-pipeline strategy. Each worker retrieves series ready to be analyzed by constantly polling the dedicated cRing (cf. \Cref{fig:dual_pipeline}). The analytics manager leverages two key components, namely $(i)$ \emph{Dynamic Batching} and $(ii)$ \emph{Approximate Caching}. 
In the remainder of this section, we further detail the design and internal architecture of batching and caching processing.



\subsection{Batching}
\label{sec:batching}

Partitioning time series into batches has two main advantages: ($i$) it reduces the communication overhead by amortizing transmission overhead, and ($ii$) it enables a faster computation through parallel processing and memory access.
According to our preliminary benchmark in \Cref{sec:rqrmnt}, the bigger the batch, the shorter the per-series processing delay \ie{} the time needed for sending, processing, and receiving the analytics results of a single series. Hence, if we only take into account processing throughput, the bigger the batch, the better. 
\def\nbseries{r}

\begin{algorithm}[t]
\footnotesize
\SetAlFnt{\small}
\SetAlgoLined
\SetKwData{Padding}{padding}
\textbf{At timeout $T$ expiration}\\
\Begin{
  $\nbseries \leftarrow$ length of cRing\;
  $B \leftarrow \min \left\{b \in \mathcal B: b\geq \nbseries\right\}$\;
  batch[$0..\nbseries$] $\leftarrow$ cRing[$0..\nbseries$]\;
  \Padding $\leftarrow$ $B- \nbseries$\;
  \If{\Padding}{
    batch[$\nbseries..B$] $\leftarrow$ pad[0..padding]\;
  }
  Process(batch)\;
 }
 \caption{Timeout based dynamic batching}
 \label{alg:timeout}
\end{algorithm}

However, big batches take longer to be filled, which can be a problem at low series arrival rates. Thus, system designers have to tradeoff analytics throughput and operational delay with respect to the series arrival rate and the application requirement. Overall, adopting a static batch size leads to systems working well only on a single operational point. Conversely, we implement in \acrshort{fix} a dynamic batching strategy with variable batch size to tradeoff analytics throughput, delay, and processing power. 

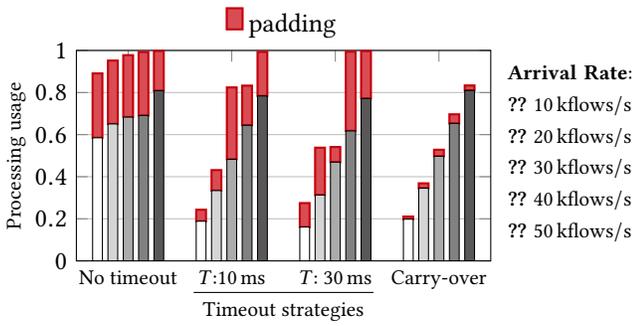
\begin{figure}[t!]
    \centering
    \def\yrat{1}
    \def\rat{0.88}
    \pgfplotscreateplotcyclelist{myCycleA}{%
  {draw=black, fill=white},
    {draw=colorRed, thick, fill=colorRed!70}
}
\pgfplotscreateplotcyclelist{myCycleB}{%
  {draw=black, fill=colorGray!25},
    {draw=colorRed, thick, fill=colorRed!70}
}
\pgfplotscreateplotcyclelist{myCycleC}{%
  {draw=black, fill=colorGray!50},
    {draw=colorRed, thick, fill=colorRed!70}
}
\pgfplotscreateplotcyclelist{myCycleD}{%
  {draw=black, fill=colorGray!75},
    {draw=colorRed, thick, fill=colorRed!70}
}
\pgfplotscreateplotcyclelist{myCycleE}{%
  {draw=black, fill=colorGray!100},
    {draw=colorRed, thick, fill=colorRed!70}
}

\def\barw{3.7pt}
\def\shift{5.8pt}

\begin{tikzpicture}
  \pgfplotsset{
    scale only axis,
    ybar stacked, 
    width = 0.65\columnwidth,
    height=\heightfigure,
    ymin=0, 
    ymax=1, 
    xmin=0.5, 
    xmax=4.5, 
    xtick=data,
    }
    
  \begin{axis}[
    bar shift=-2*\shift,
    bar width = \barw,
    ylabel={Processing usage},
    label style = {
        font=\footnotesize,
        inner sep=0pt,
    },
    legend style = {
        anchor= south,
        at = {(0.5, 1.02)},
        draw=none,
        },
    x tick label style = {
        font = \footnotesize},
    xticklabels={
        {No timeout},
        {$T$:\SI{10}{ms}},
        {$T$: \SI{30}{ms}},
        {\carry}
    },
    ymajorgrids,
    cycle list name = myCycleA
    ]
    
    \addplot+ [] table 
        [x index = 0, y index = 2] {serieA.dat};         
        \label{ten}
    \addplot+ [] table 
        [x index = 0, y index = 1] {serieA.dat};

    \legend{, padding}
  \end{axis}
 
  
  \begin{axis}[
    axis line style={draw=none},
    tick style={draw=none},
    yticklabels=\empty,
    xticklabels=\empty,    
    bar shift = -1*\shift,
    bar width = \barw,
    cycle list name = myCycleB
    ]
    \addplot+ [] table
        [x index = 0, y index = 2] {serieB.dat};
        \label{twenty}
    \addplot+ [] table
        [x index = 0, y index = 1] {serieB.dat};

  \end{axis}
  
  
  \begin{axis}[
    axis line style={draw=none},
    tick style={draw=none},
    bar shift = 0*\shift,
    bar width = \barw,
    yticklabels=\empty,
    xticklabels=\empty,    
    cycle list name = myCycleC
    ]
    \addplot+ [] table
        [x index = 0, y index = 2] {serieC.dat};
        \label{thirty}    
    \addplot+ [] table
        [x index = 0, y index = 1] {serieC.dat};
  \end{axis}
  
 
 \begin{axis}[
    axis line style={draw=none},
    tick style={draw=none},
    bar shift = 1*\shift,
    bar width = \barw,
    yticklabels=\empty,
    xticklabels=\empty,    
    cycle list name = myCycleD
    ]
    \addplot+ [] table
        [x index = 0, y index = 2] {serieD.dat};
        \label{fourty}    
    \addplot+ [] table
        [x index = 0, y index = 1] {serieD.dat};
  \end{axis}
  

 \begin{axis}[
    axis line style={draw=none},
    tick style={draw=none},
    bar shift = 2*\shift,
    bar width = \barw,
    yticklabels=\empty,
    xticklabels=\empty,    
    cycle list name = myCycleE
    ]
    \addplot+ [] table
        [x index = 0, y index = 2] {serieE.dat};
        \label{fifty}    
    \addplot+ [] table 
        [x index = 0, y index = 1] {serieE.dat};
  \end{axis}

\node [
    anchor = north west,
    draw=none,
    fill=white,
    text width = 10em,
    font = \footnotesize
    ] 
    at (rel axis cs: 1.02,0.98) 
    {\baselineskip=12pt\textbf{Arrival Rate}:\\
    \ref{ten} \SI{10}{kflows/s}\\
    \ref{twenty} \SI{20}{kflows/s}\\
    \ref{thirty} \SI{30}{kflows/s}\\
    \ref{fourty} \SI{40}{kflows/s}\\
    \ref{fifty} \SI{50}{kflows/s}\\};
  
\node [
    anchor = north,
    draw=none,
    fill=white,
    font = \footnotesize
    ] 
    at (rel axis cs: 0.5, -0.15)
    (timeout)
    {Timeout strategies};
    
\draw 
    (timeout.north west) to
    (timeout.north east);

\end{tikzpicture}
    \vspace{-0.3cm}
    \caption{Processing usage for TPU (\num{1} core). The \carry{} strategy is set with timeout at \SI{10}{\milli\second} and $\Phi=\num{0.2}$.}
    \label{fig:dummies_processing}
    \vspace{-0.3cm}
\end{figure}

The concept of \emph{dynamic batch} consists in running the \gls{dl} model with variable batch size~\cite{clipper}. Due to the limited amount of available resources in \gls{dl} accelerators, a \gls{dl} model cannot be used with any batch size; it is defined with one unique batch size $B$. Dynamic batching is then implemented by hosting multiple models, each one with a different batch size as in~\cite{triton,mindstudio}. A naive approach to dynamic batching is presented in \cref{alg:timeout}. Let $\mathcal B$ be the set of batch sizes of the implemented \gls{dl} models. When the system schedules models inference (\eg{} using a timeout), $\nbseries$ series are waiting in the cRing. The Analytics manager creates a new batch from these $\nbseries$ series by selecting the model with batch size $B$ such that $B$ is the smallest batch size greater than $\nbseries$ in $\mathcal B$. If the batch is not complete, the system may add $B-\nbseries$ padding series before sending it to the hardware accelerator.

To better understand the behavior of such an algorithm in realistic network conditions, we simulated system behaviors by extracting single-chip TPU's performance profile reported in \Cref{sec:rqrmnt} for batch sizes equal to $2^{x}$ with $x \in [3..10]$. The simulation input is then generated by a Poisson process that inputs a flow randomly chosen from flows extracted from the \textit{campus}, \textit{home}, and \textit{transit} datasets at a given average series arrival rate $\lambda$. \Cref{fig:dummies_processing} presents the total hardware accelerator usage in the simulated scenario, which is further split in padding and series with different timeout values \ie{} left, no timeout, and center fixed timeout. Notice that we here consider timeouts that are in the order of magnitude of $(i)$ the RTT to get analytics as soon as possible and $(ii)$ the analytics processing delay reported in \Cref{sec:rqrmnt}. 

When no timeout applies, the system sends a new batch immediately when the accelerator is ready. In this case, the processing usage is close to \num{1} even when the series arrival rate is low. This strategy is inefficient in two aspects: $(i)$ at low arrival rates, the number of series $\nbseries$ is low, so the system should add a large number of padding, which is a waste of resources; $(ii)$ since the number of series $\nbseries$ is low, the size of the batch $B$ is also low, which is less efficient in terms of processing throughput. The timeout addresses the problem of over-utilization of the accelerator since the hardware is used only a fraction of time at low series arrival rates. We observe however that a large amount of computational power is spent on processing padding, regardless of the timeout.

To prevent the accelerator to uselessly process padding, we design a custom dynamic batching system. 
\acrshort{fix} adopts both concepts of dynamic batching and timeout $T$ because they provide an efficient basis for trade-off throughput and application delay. Note that since the cRing is pulled in pool mode a real timeout does not exist, we nonetheless present the batching strategy with a real timeout for ease of presentation. We augment the timeout batching strategy with a \emph{\carry{}} system to control the padding as illustrated in \Cref{alg:padding}. Let $\Phi$ be a threshold in $\left [ 0, 0.5\right ]$ that sets the maximum fraction of padding accepted in the batch. When the timeout is reached, the system contains $\nbseries$ series. Let $B$ be the smallest batch size in $\mathcal B$ greater than $\nbseries$. If the padding necessary to complete the batch is greater than $\Phi$, the system scales back to a smaller batch size $B'$, which is the largest batch size in $\mathcal B$ smaller than $\nbseries$. This way, we do not include in the batch the latest series that arrived in the series ring. These series wait for the next batch that will be processed. 

In essence, the \carry{} policy $(i)$ adds some extra delay for a subset of series, but these series have just arrived in the system so the extra delay compared to the timeout is small; and $(ii)$ enables control of the fraction of padding in the system. The rightmost part of \Cref{fig:dummies_processing} presents the results in the simulated scenario for the \carry{} policy with $\Phi=0.2$ and timeout = \SI{10}{ms} demonstrating that it is able to control both padding and processing efficiency with respect to no timeout and fixed timeout batching policies. We further analyze the performance of our batching strategy in \Cref{sec:am_bench} using the \acrshort{fix} prototype.

\begin{algorithm}[t]
\footnotesize
\SetAlgoLined
\SetKwData{Padding}{padding}
\SetKwData{P}{$\Phi$}


\textbf{At timeout $T$ expiration}\\
\Begin{
  $\nbseries \leftarrow$ length of cRing\;
  $B \leftarrow \min \left\{b \in \mathcal B: b\geq \nbseries\right\}$\;
  \Padding $\leftarrow$ $B-\nbseries$\;
  \eIf{$\frac{\Padding}{B}$ > \P}{
    $B \leftarrow \max \left\{b \in \mathcal B: b\leq \nbseries\right\}$\;
    batch[$0..B$] $\leftarrow$ cRing[$0..B$]\;
  }
  {
    batch[$0..\nbseries$] $\leftarrow$ cRing[$0..\nbseries$]\;
    \If{\Padding}{
      batch[$\nbseries..B$] $\leftarrow$ pad[0..\Padding]\;
    }
  }
  Process(batch)\;
 }
 \caption{\carry{} dynamic batching}
 \label{alg:padding}
\end{algorithm}


\subsection{Caching}\label{sec:caching}

\def\nbseries{1534}
\def\nbflows{8946}
\def\nbapps{168}
\def\nbcats{5}

\begin{algorithm}[t]
\footnotesize
\SetAlgoLined
\SetKwData{Sdelta}{$S_\delta$}
\SetKwData{cRing}{cRing}
\SetKwData{B}{B}
\SetKwData{lookup}{$l_{S_K}$ $\leftarrow$ Cache.lookup($S_\delta$)}

\textbf{At timeout $T$ expiration}\\
\Begin{
  \For{$z\leftarrow 0$ \KwTo len(\cRing)}{
    $S_K$ $\leftarrow$ \cRing[z]\;
    $S_\delta$ $\leftarrow$ $s_{1\leq i \leq \delta} \in S_K$\;
  \If{\lookup}{
    pull(cRing[z])\;
  }
 }
  batchComposition()\;
 }
 \smallskip
 \textbf{At analytics result reception}\\
 \Begin{
   \For{$z\leftarrow 0$ \KwTo \B}{
     $S_K$ $\leftarrow$ batch[z]\;    
     $S_\delta$ $\leftarrow$ $s_{1\leq i \leq \delta} \in S_K$\;
     Cache.insert($S_\delta$)\;
   }
 }
 \caption{Prefix cache}
 \label{alg:cache}
\end{algorithm}

A trained \gls{dl} model is deterministic. In the case of traffic classification it can be abstracted as a non-linear function $f(\cdot)$ that maps a series $S \in \mathcal S_K$, \ie{} with features extracted from the first $K$ packets, to a class, \ie{} label $l \in \mathcal L$. The label set $\mathcal L$ depends on the analytics. In the traffic classification example, we can classify 200 different applications. 
The function $f$ is not injective: multiple series can have the same label $l \in \mathcal L$. Based on this observation and on the fact that multiple flows have the same series (see \Cref{tab:dataset}), we foresee the benefits of implementing a cache, which stores popular analytics computation, to both speed up analytics and reduce the load on the hardware accelerator. 

In our context, the caching system $\mathcal C$ stores $C$ entries in the form of keys and values, where the key is a series $S_K \in \mathcal S_K$ and the value is an associated label $l_{S_K}$. An incoming flow for which the extracted series $S_K$ is cached in $\mathcal C$ is directly classified with the stored label. The performance of such cache, in terms of hit ratio increases when $(i)$ the number of different flows having the same series is large and $(ii)$ the distribution of flows per series (popularity) is skewed. 




To increase the hit-ratio of the cache, a cache designer can implement \emph{approximate caching} (\aka{} similarity caching)~\cite{PandeyBCJKV09,MiguelAMJ15,FalchiLOPR12} by $(i)$ reducing the set of keys $\mathcal P$ in the cache and $(ii)$ applying an approximate function that maps the set of input series $\mathcal S_K$ to a smaller set of keys $\mathcal P'$.   Without loss of generality,  we focus in the context of our use case on \emph{prefix caching}, where any cached key is a subset of the series $S_K$, \ie{} the key set $\mathcal P'$ is the subset $\mathcal S_\delta$ where $\delta < K$.
Formally, we define a function $q_\delta(\cdot)$ that transforms a series in $S_K \in \mathcal S_K$ to a series $S_\delta \in \mathcal S_\delta$ where, for any $1\leq i \leq \delta$, and for any series' feature $s' \in S_K$ with $s' = q_\delta(s) \in S_\delta$, $s'_i = s_i$. 
Notice that other time-series based analytics (e.g., forecast of load or other signals) would equally benefit from policies with very similar implementation (e.g., \emph{postfix caching}, to give more importance to most recent samples), so while the quantitative evaluation is limited to the traffic classification use-case, the qualitative lessons holds to a larger extent.




\acrshort{fix} implements the \textit{prefix cache} as a \gls{lru} cache that act as filter before the batch composition described in \Cref{alg:padding}. Note that in the real implementation caching and batching modules are entangled, but we present them separately for the sake of clarity. The process at timeout expiration and at analytics result reception is described in \Cref{alg:cache}. When the timeout expires, before the batch composition and for each series in the cRing, the approximation function $q_\delta(\cdot)$ is applied to series $S_K$ to obtain the prefix series $S_\delta$ for which a cache lookup is performed. In case $S_\delta$ has a matching label $l$, \ie{} \textit{Hit}, \acrshort{fix} tags the series $S_K$ with the label $l$ and pulls the element from the cRing since the analytics is considered as already executed. Notice that, according to the cache replacement policy, \textit{Cache.lookup} also adjusts internal cache data structures \eg{} updates the list of least recently used elements for \gls{lru}. In case $S_\delta$ does not have a matching label $l$, \ie{} \textit{Miss}, the processing continues. Finally, 
a batch is composed using \Cref{alg:padding}. At analytics results reception, \acrshort{fix} inserts the label $l$ in the cache for the corresponding prefix series $S_\delta$ derived from $S_K$.

\def\pref{6}
\begin{figure}
    \centering
    \def\yrat{1}
    \def\rat{0.65}
    \def\widthbar{6pt}
\pgfplotscreateplotcyclelist{mySpecialCycle}{%
  {colorRed, thick, 
    pattern=mydots,
    pattern color=colorRed},
  {colorRed, thick, 
    pattern=mydots,
    pattern color=colorRed!30},
}
\pgfplotscreateplotcyclelist{mySpecialCycleGreen}{%
  {colorGreen, thick, fill = colorGreen!80},
  {colorGreen, thick, fill=colorGreen!30}
}

\pgfplotscreateplotcyclelist{mySpecialCycleBlue}{%
  {black, thick,
  pattern=crosshatch,
  pattern color = colorGray}
}

\def\xmin{45}
\def\xmax{325}
\def\ymax{0.5}
\pgfplotstableread[col sep = comma]
    {Data/hit_error_adjusted_size.csv}
    \mytable

\begin{tikzpicture}
    \begin{axis}[
        ymin = 0, ymax = \ymax,
        xmin = \xmin, xmax = \xmax,
        scaled y ticks=false,
        scaled x ticks=false,
        ylabel near ticks,
        scale only axis,
        ybar stacked,
        bar width = \widthbar,
        width = \rat*\columnwidth, 
        height = \yrat*\heightfigure,
        xlabel = Cache unit size, 
        ylabel = Hit ratio,
        yticklabel style={
            font = \footnotesize
        },
        xticklabel style={
            font = \footnotesize
        },
        label style = {
            font=\small
        },
        ylabel style = {
            font = \small,
            inner sep = 1pt,
        },
        legend style={
            font=\scriptsize,
            at={(1.02, 1.)},
            anchor = north west,
            draw = none,
            inner sep=1pt,
        },
        legend columns=1,
        legend cell align={left},
        ymajorgrids,
        xmajorgrids,
        yminorgrids,
        enlarge x limits=0.03,
        enlarge y limits = 0.01,
        cycle list name=mySpecialCycle
    ]
    
    \addlegendimage{empty legend}
    \addlegendentry{\textbf{Prefix 4}}
        \addplot+ [
            ] table [
                x = cache_size, 
                y expr= \thisrow{hit_4} - \thisrow{app_4}
            ] {\mytable};
            \addlegendentry{good}
        \addplot+ [] table [
                x = cache_size, 
                y expr= \thisrow{app_4}
            ] {\mytable};
            \addlegendentry{error}
    \end{axis}
    

    \begin{axis}[
        ymin = 0, ymax = \ymax,
        xmin = \xmin, xmax = \xmax,
        axis x line=none,
        axis y line=none,
        scale only axis,
        ybar stacked,
        bar width = \widthbar,
        width = \rat*\columnwidth, 
        height = \yrat*\heightfigure,
        legend style={
            font=\scriptsize,
            at={(1.02, 0.5)},
            anchor = north west,
            draw = none,
            inner sep=1pt,
        },
        legend columns=1,
        legend cell align={left},
        enlarge x limits=0.02,
        enlarge y limits = 0,
        cycle list name=mySpecialCycleGreen
    ]
    
    \addlegendimage{empty legend}
    \addlegendentry{\textbf{Prefix \pref}}

    \addplot+ [
            xshift=0.9*\widthbar,
            legend image post style={xshift=-0.9*\widthbar}] 
        table [
                x = cache_size, 
                y expr= \thisrow{hit_\pref} - \thisrow{app_\pref}
            ] {\mytable};
    \addlegendentry{good}
    
    \addplot+ [
            xshift=0.9*\widthbar,
            legend image post style={xshift=-0.9*\widthbar}] 
        table [
                x = cache_size, 
                y expr= \thisrow{app_\pref}
            ] {\mytable};
    \addlegendentry{error}
    \end{axis}
    
  \begin{axis}[
        ymin = 0, ymax = \ymax,
        xmin = \xmin, xmax = \xmax,
        axis x line=none,
        axis y line=none,
        scale only axis,
        ybar stacked,
        bar width = \widthbar,
        width = \rat*\columnwidth, 
        height = \yrat*\heightfigure,
        legend style={
            font=\scriptsize,
            at={(1.02, 0.)},
            draw = none,
            anchor = north west,
            inner sep=1pt,
        },
        legend columns=1,
        legend cell align={left},
        enlarge x limits=0.02,
        enlarge y limits = 0,
        cycle list name=mySpecialCycleBlue
    ]
    \addlegendimage{empty legend}
    \addlegendentry{\textbf{Regular cache}}
    \addplot+ [
            black, thick,
            pattern=crosshatch,
            pattern color = colorGray,
            xshift=1.8*\widthbar,
            legend image post style={xshift=-1.8*\widthbar}] 
        table [
                x = cache_size, 
                y expr= \thisrow{hit_10}%
            ] {\mytable};
    \addlegendentry{good}
    \end{axis}

\end{tikzpicture}
    \caption{Prefix cache hit-ratio for prefixes $\delta=4$ and $\delta=\pref$, in comparison to regular cache with \num{10} packets.}
    \label{fig:hit-ratio}
    \vspace{-0.3cm}
\end{figure}
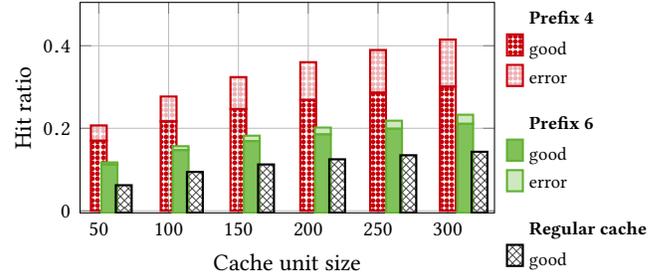

\begin{figure*}[!ht]
\centering
\def\ratio{0.31} 
\def\xrat{0.45} 
\def\yrat{0.9} 
\def\barw{0.15}
\def\ratw{3.9em}
\pgfplotscreateplotcyclelist{uniformCyclelist}{%
  {draw=colorRed, pattern=mydots, pattern color=colorRed},
  {draw=black, fill=colorGray!70, 
    },
  {draw=colorGray, fill=colorGray!40
  },
  {draw=colorGray!60, fill=colorGray!20 
  }
}
\subcaptionbox{Delay to obtain the label.\label{fig:delay_carryon}}
  [\ratio\textwidth] 
  {\def\rat{0.03}

\begin{tikzpicture}
    \begin{axis}[
        scale only axis,
        boxplot/draw direction=y,
        boxplot = {
            draw position={1/5 + floor(\plotnumofactualtype/4) + 1/5*mod(\plotnumofactualtype,4)},
            box extend=\barw,
        },
        ymin = 0, 
        xmin=0, xmax=3,
        width= \xrat*\columnwidth, 
        height = \yrat*\heightfigure,
        ylabel = {Delay from \ith{10} packet to label [ms]},
        ylabel near ticks,
        ylabel style = {
            align = center,
            font=\scriptsize,
            text width = \yrat*\heightfigure,
        },
        x=\ratw,
        xticklabels={
            {0},
            {5ms},%
            {10ms},%
            {20ms}%
        },
        x tick label style={
            text width=0.05*\columnwidth,
            align=center,
            font = \footnotesize
        },
        xtick distance=1,
        x tick label as interval,
        y tick label style={
            font=\footnotesize,
        },
        legend style={
            font = \footnotesize,
            draw = none,
            fill = none,
            inner sep = 1pt,
            at = {(0.8,1.02)},
            anchor = south west,
            /tikz/every even column/.append style={
                column sep=0.5cm}
        },
        legend columns=4,
        legend cell align={left},
        boxplot legend,
        ymajorgrids,
        enlargelimits=0.01,
        cycle list name=uniformCyclelist
    ]

    \pgfplotstableread[col sep = comma]
    {Data/parsed_ttlabel.csv}
    \mytable

    \pgfplotstablegetrowsof{\mytable}
    \pgfmathsetmacro\numberofrows{\pgfplotsretval-1}

    \pgfplotsinvokeforeach{0,...,\numberofrows}{
        \addplot+[
            boxplot prepared from table={
                table=\mytable,
                row =#1,
                lower whisker=q01_fixed,
                upper whisker=q99_fixed,
                lower quartile=q25_fixed,
                upper quartile=q75_fixed,
                median=q50_fixed
                }, 
            boxplot prepared]
        coordinates {};
        
        \addplot+[
            boxplot prepared from table={
                table=\mytable,
                row =#1,
                lower whisker=q01_carryon_0.1,
                upper whisker=q99_carryon_0.1,
                lower quartile=q25_carryon_0.1,
                upper quartile=q75_carryon_0.1,
                median=q50_carryon_0.1
                }, 
            boxplot prepared]
        coordinates {};

        \addplot+[
            boxplot prepared from table={
                table=\mytable,
                row =#1,
                lower whisker=q01_carryon_0.2,
                upper whisker=q99_carryon_0.2,
                lower quartile=q25_carryon_0.2,
                upper quartile=q75_carryon_0.2,
                median=q50_carryon_0.2
                }, 
            boxplot prepared]
        coordinates {};

        \addplot+[
            boxplot prepared from table={
                table=\mytable,
                row =#1,
                lower whisker=q01_carryon_0.3,
                upper whisker=q99_carryon_0.3,
                lower quartile=q25_carryon_0.3,
                median=q50_carryon_0.3
                }, 
            boxplot prepared]
        coordinates {};
    }
    \legend{
        Timeout, 
        \carry{} $\Phi=0.1$,
        \carry{} $\Phi=0.2$,
        \carry{} $\Phi=0.3$
    }
    \end{axis}
\end{tikzpicture}}
\hfill%
\subcaptionbox{Ratio of padding.\label{fig:padding_carryon}}
  [\ratio\textwidth] 
  {\def\rat{0.03}
\begin{tikzpicture}
    \begin{axis}[
        scale only axis,
        boxplot/draw direction=y,
        boxplot = {
            draw position={1/5 + floor(\plotnumofactualtype/4) + 1/5*mod(\plotnumofactualtype,4)},
            box extend=\barw,
        },
        ymin = 0, ymax=0.5,
        xmin=0, xmax=3,
        width= \xrat*\columnwidth, 
        height = \yrat*\heightfigure,
        ylabel = Per-batch padding ratio,
        ylabel near ticks,
        ylabel style = {
            align = center,
            font=\scriptsize,
            text width = \yrat*\heightfigure,
        },
        x=\ratw,
        xticklabels={
            {0},
            {5ms},%
            {10ms},%
            {20ms}%
        },
        x tick label style={
            text width=0.02*\columnwidth,
            align=center,
            font = \footnotesize
        },
        xtick distance=1,
        x tick label as interval,
        y tick label style={
            font=\footnotesize,
        },
        ymajorgrids,
        enlargelimits=0.01,
        cycle list name=uniformCyclelist
    ]

    \pgfplotstableread[col sep = comma]
    {Data/parsed_padding.csv}
    \mytable

    \pgfplotstablegetrowsof{\mytable}
    \pgfmathsetmacro\numberofrows{\pgfplotsretval-1}

    \pgfplotsinvokeforeach{0,...,\numberofrows}{
        \addplot+[
            boxplot prepared from table={
                table=\mytable,
                row =#1,
                lower whisker=empty,
                upper whisker=q99_fixed,
                lower quartile=empty,
                upper quartile=q50_fixed,
                median=q50_fixed
                }, 
            boxplot prepared]
        coordinates {};
        
        \addplot+[
            boxplot prepared from table={
                table=\mytable,
                row =#1,
                lower whisker=empty,
                upper whisker=q99_carryon_0.1,
                lower quartile=empty,
                upper quartile=q50_carryon_0.1,
                median=q50_carryon_0.1
                }, 
            boxplot prepared]
        coordinates {};

        \addplot+[
            boxplot prepared from table={
                table=\mytable,
                row =#1,
                lower whisker=empty,
                upper whisker=q99_carryon_0.2,
                lower quartile=empty,
                upper quartile=q50_carryon_0.2,
                median=q50_carryon_0.2
                }, 
            boxplot prepared]
        coordinates {};

        \addplot+[
            boxplot prepared from table={
                table=\mytable,
                row =#1,
                lower whisker=empty,
                upper whisker=q99_carryon_0.3,
                lower quartile=empty,
                upper quartile=q50_carryon_0.3,
                median=q50_carryon_0.3
                }, 
            boxplot prepared]
        coordinates {};
    }%
    \end{axis}
\end{tikzpicture}}
 \hfill%
\subcaptionbox{Ratio of post-mortem.\label{fig:dead_carryon}}
  [\ratio\textwidth] 
  {\def\rat{0.03}
\begin{tikzpicture}
    \begin{axis}[
        scale only axis,
        boxplot/draw direction=y,
        boxplot = {
            draw position={1/5 + floor(\plotnumofactualtype/4) + 1/5*mod(\plotnumofactualtype,4)},
            box extend=\barw,
        },
        ymin = 0, ymax=0.2,
        xmin=0, xmax=3,
        width= \xrat*\columnwidth, 
        height = \yrat*\heightfigure,
        ylabel = Ratio of post-mortem analytics,
        ylabel near ticks,
        ylabel style = {
            align = center,
            font=\scriptsize,
            text width = \yrat*\heightfigure,
        },
        ytick={0, 0.1, 0.2},
        x=\ratw,
        xticklabels={
            {0},
            {5ms},%
            {10ms},%
            {20ms}%
        },
        x tick label style={
            text width=0.02*\columnwidth,
            align=center,
            font = \footnotesize
        },
        xtick distance=1,
        x tick label as interval,
        y tick label style={
            font=\footnotesize,
        },
        ymajorgrids,
        enlargelimits=0.01,
        cycle list name=uniformCyclelist
    ]

    \pgfplotstableread[col sep = comma]
    {Data/parsed_dead.csv}
    \mytable

    \pgfplotstablegetrowsof{\mytable}
    \pgfmathsetmacro\numberofrows{\pgfplotsretval-1}

    \pgfplotsinvokeforeach{0,...,\numberofrows}{
        \addplot+[
            boxplot prepared from table={
                table=\mytable,
                row =#1,
                lower whisker=empty,
                upper whisker=q99_fixed,
                lower quartile=empty,
                upper quartile=q50_fixed,
                median=q50_fixed
                }, 
            boxplot prepared]
        coordinates {};
        
        \addplot+[
            boxplot prepared from table={
                table=\mytable,
                row =#1,
                lower whisker=empty,
                upper whisker=q99_carryon_0.1,
                lower quartile=empty,
                upper quartile=q50_carryon_0.1,
                median=q50_carryon_0.1
                }, 
            boxplot prepared]
        coordinates {};

        \addplot+[
            boxplot prepared from table={
                table=\mytable,
                row =#1,
                lower whisker=empty,
                upper whisker=q99_carryon_0.2,
                lower quartile=empty,
                upper quartile=q50_carryon_0.2,
                median=q50_carryon_0.2
                }, 
            boxplot prepared]
        coordinates {};

        \addplot+[
            boxplot prepared from table={
                table=\mytable,
                row =#1,
                lower whisker=empty,
                upper whisker=q99_carryon_0.3,
                lower quartile=empty,
                upper quartile=q50_carryon_0.3,
                median=q50_carryon_0.3
                }, 
            boxplot prepared]
        coordinates {};
    }%
    \end{axis}
\end{tikzpicture}}
\caption{Timeout and Carry-over batching policy. Boxplots show \ith{99}, \ith{75}, \ith{25}, and $1^{\text{st}}$ percentiles of measures while barcharts report median and \ith{99} percentile.}
\label{fig:carryon}
\end{figure*}
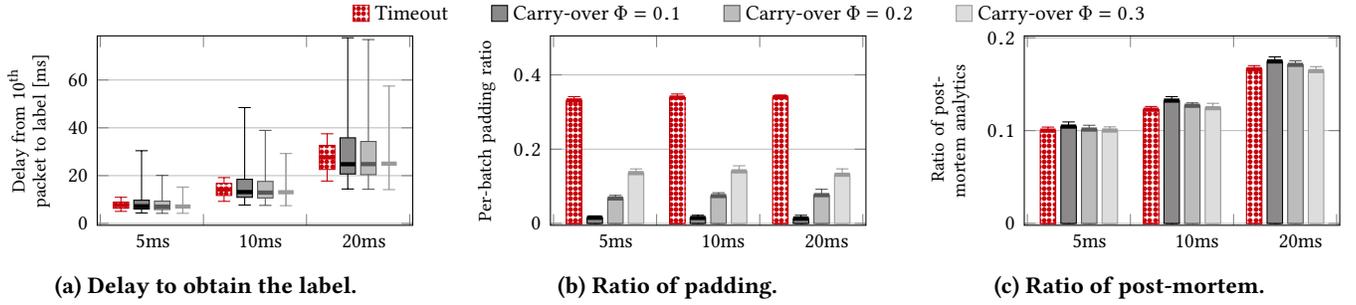

\begin{figure}[!ht]
\centering
\def\ratio{0.31} 
\def\xrat{0.45} 
\def\yrat{0.9} 
\def\barw{0.14}
\def\ratw{3.6em}
\pgfplotscreateplotcyclelist{uniformCyclelist}{%
  {colorRed, fill=colorRed!50},
  {colorGray, fill=colorGray!50
  },
  {colorRed, pattern=mydots,
  pattern color = colorRed
  },
  {colorGray, pattern=mydots, pattern color=colorGray}
}
\subcaptionbox{Delay to label\label{fig:delay_cache}}
  [\ratio\columnwidth] 
  {\def\rat{0.03}

\begin{tikzpicture}
    \begin{axis}[
        scale only axis,
        boxplot/draw direction=y,
        boxplot = {
            draw position={1/5 + floor(\plotnumofactualtype/4) + 1/5*mod(\plotnumofactualtype,4)},
            box extend=\barw,
        },
        ymin = 0, 
        xmin=0, xmax=1,
        width= \xrat*\columnwidth, 
        height = \yrat*\heightfigure,
        ylabel = {Delay from \ith{10} packet to label [ms]},
        ylabel style = {
            align = center,
            font=\scriptsize,
            text width = \yrat*\heightfigure,
            at={(axis description cs:-0.25,.5)}},
        x=\ratw,
        xticklabels={
            {0},
        },
        x tick label style={
            text width=0.02*\columnwidth,
            align=center,
            font = \footnotesize
        },
        xtick distance=1,
        x tick label as interval,
        y tick label style={
            font=\footnotesize,
        },
        legend style={
            font = \scriptsize,
            draw = none,
            fill = none,
            inner sep = 1pt,
            at = {(-0.65,1.02)},
            anchor = south west,
            /tikz/every even column/.append style={
                column sep=0.22cm}
        },
        legend columns=4,
        legend cell align={left},
        boxplot legend,
        ymajorgrids,
        enlargelimits=0.01,
        cycle list name=uniformCyclelist
    ]

    \pgfplotstableread[col sep = comma]
    {Data/parsed_ttlabel_second.csv}
    \mytable

    \pgfplotstablegetrowsof{\mytable}
    \pgfmathsetmacro\numberofrows{\pgfplotsretval-1}

   \pgfplotsinvokeforeach{0,...,\numberofrows}{
        \addplot+[
            boxplot prepared from table={
                table=\mytable,
                row =#1,
                lower whisker=q01_fixed,
                upper whisker=q99_fixed,
                lower quartile=q25_fixed,
                upper quartile=q75_fixed,
                median=q50_fixed
                }, 
            boxplot prepared]
        coordinates {};
        
        \addplot+[
            boxplot prepared from table={
                table=\mytable,
                row =#1,
                lower whisker=q01_carryon,
                upper whisker=q99_carryon,
                lower quartile=q25_carryon,
                upper quartile=q75_carryon,
                median=q50_carryon
                }, 
            boxplot prepared]
        coordinates {};

        \addplot+[
            boxplot prepared from table={
                table=\mytable,
                row =#1,
                lower whisker=q01_fixed-identity,
                upper whisker=q99_fixed-identity,
                lower quartile=q25_fixed-identity,
                upper quartile=q75_fixed-identity,
                median=q50_fixed-identity
                }, 
            boxplot prepared]
        coordinates {};

        \addplot+[
            boxplot prepared from table={
                table=\mytable,
                row =#1,
                lower whisker=q01_carryon-identity,
                upper whisker=q99_carryon-identity,
                lower quartile=q25_carryon-identity,
                upper quartile=q75_carryon-identity,
                median=q50_carryon-identity
                }, 
            boxplot prepared]
        coordinates {};
    }%
    \legend{
        Timeout, 
        \carry,
        Timeout + caching,
        \carry + caching
    }
    \end{axis}
\end{tikzpicture}}
\hfill%
\subcaptionbox{Ratio of padding\label{fig:padding_cache}}
  [\ratio\columnwidth] 
  {\def\rat{0.03}
\begin{tikzpicture}
    \begin{axis}[
        scale only axis,
        boxplot/draw direction=y,
        boxplot = {
            draw position={1/5 + floor(\plotnumofactualtype/4) + 1/5*mod(\plotnumofactualtype,4)},
            box extend=\barw,
        },
        ymin = 0, ymax=0.5,
        xmin=0, xmax=1,
        width= \xrat*\columnwidth, 
        height = \yrat*\heightfigure,
        ylabel = Per-batch padding ratio,
        ylabel style = {
            align = center,
            font=\scriptsize,
            text width = \yrat*\heightfigure,
            at={(axis description cs:-0.30,.5)}
        },
        x=\ratw,
        xticklabels={
            {0},
        },
        x tick label style={
            text width=0.02*\columnwidth,
            align=center,
            font = \footnotesize
        },
        xtick distance=1,
        x tick label as interval,
        y tick label style={
            font=\footnotesize,
        },
        ymajorgrids,
        enlargelimits=0.01,
        cycle list name=uniformCyclelist
    ]

    \pgfplotstableread[col sep = comma]
    {Data/parsed_padding_second.csv}
    \mytable

    \pgfplotstablegetrowsof{\mytable}
    \pgfmathsetmacro\numberofrows{\pgfplotsretval-1}

    \pgfplotsinvokeforeach{0,...,\numberofrows}{
        \addplot+[
            boxplot prepared from table={
                table=\mytable,
                row =#1,
                lower whisker=empty,
                upper whisker=q99_fixed,
                lower quartile=empty,
                upper quartile=q50_fixed,
                median=q50_fixed
                }, 
            boxplot prepared]
        coordinates {};
        
        \addplot+[
            boxplot prepared from table={
                table=\mytable,
                row =#1,
                lower whisker=empty,
                upper whisker=q99_carryon,
                lower quartile=empty,
                upper quartile=q50_carryon,
                median=q50_carryon
                }, 
            boxplot prepared]
        coordinates {};

        \addplot+[
            boxplot prepared from table={
                table=\mytable,
                row =#1,
                lower whisker=empty,
                upper whisker=q99_fixed-identity,
                lower quartile=empty,
                upper quartile=q50_fixed-identity,
                median=q50_fixed-identity
                }, 
            boxplot prepared]
        coordinates {};

        \addplot+[
            boxplot prepared from table={
                table=\mytable,
                row =#1,
                lower whisker=empty,
                upper whisker=q99_carryon-identity,
                lower quartile=empty,
                upper quartile=q50_carryon-identity,
                median=q50_carryon-identity
                }, 
            boxplot prepared]
        coordinates {};
    }%
    \end{axis}
\end{tikzpicture}}
 \hfill%
\subcaptionbox{Post-mortem ratio\label{fig:dead_cache}}
  [0.34\columnwidth] 
  {\def\rat{0.03}
\begin{tikzpicture}
    \begin{axis}[
        scale only axis,
        boxplot/draw direction=y,
        boxplot = {
            draw position={1/5 + floor(\plotnumofactualtype/4) + 1/5*mod(\plotnumofactualtype,4)},
            box extend=\barw,
        },
        ymin = 0, ymax=0.2,
        xmin=0, xmax=1,
        width= \xrat*\columnwidth, 
        height = \yrat*\heightfigure,
        ylabel = Ratio of post-mortem labeling,
        ylabel style = {
            align = center,
            font=\scriptsize,
            text width = \yrat*\heightfigure,
            at={(axis description cs:-0.28,.5)}},
        ytick={0, 0.1, 0.2},
        x=\ratw,
        xticklabels={
            {0},
        },
        x tick label style={
            text width=0.02*\columnwidth,
            align=center,
            font = \footnotesize
        },
        xtick distance=1,
        x tick label as interval,
        y tick label style={
            font=\footnotesize,
        },
        ymajorgrids,
        enlargelimits=0.01,
        cycle list name=uniformCyclelist
    ]

    \pgfplotstableread[col sep = comma]
    {Data/parsed_dead_second.csv}
    \mytable

    \pgfplotstablegetrowsof{\mytable}
    \pgfmathsetmacro\numberofrows{\pgfplotsretval-1}

    \pgfplotsinvokeforeach{0,...,\numberofrows}{
        \addplot+[
            boxplot prepared from table={
                table=\mytable,
                row =#1,
                lower whisker=empty,
                upper whisker=q99_fixed,
                lower quartile=empty,
                upper quartile=q50_fixed,
                median=q50_fixed
                }, 
            boxplot prepared]
        coordinates {};
        
        \addplot+[
            boxplot prepared from table={
                table=\mytable,
                row =#1,
                lower whisker=empty,
                upper whisker=q99_carryon,
                lower quartile=empty,
                upper quartile=q50_carryon,
                median=q50_carryon
                }, 
            boxplot prepared]
        coordinates {};

        \addplot+[
            boxplot prepared from table={
                table=\mytable,
                row =#1,
                lower whisker=empty,
                upper whisker=q99_fixed-identity,
                lower quartile=empty,
                upper quartile=q50_fixed-identity,
                median=q50_fixed-identity
                }, 
            boxplot prepared]
        coordinates {};

        \addplot+[
            boxplot prepared from table={
                table=\mytable,
                row =#1,
                lower whisker=empty,
                upper whisker=q99_carryon-identity,
                lower quartile=empty,
                upper quartile=q50_carryon-identity,
                median=q50_carryon-identity
                }, 
            boxplot prepared]
        coordinates {};
    }%
    \end{axis}
\end{tikzpicture}}
\caption{Caching impact ($\Phi=0.2, timeout=10ms)$. Boxplots show \ith{99}, \ith{75}, \ith{25}, and $1^{\text{st}}$ percentiles; barcharts report median and \ith{99} percentile.}
\vspace{-0.3cm}
\label{fig:withcache}
\end{figure}
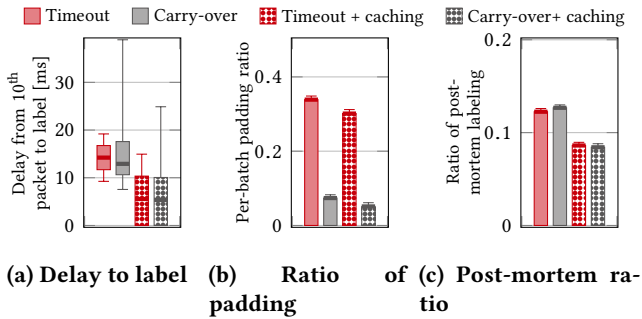

For a given hit-ratio target, the prefix cache is smaller than the regular cache. It is however an approximate cache in the sense that it is not guaranteed that the series $S_K$ and $S_\delta$ have the same analytics result. In case of a hit, two cases can be distinguished:
 either (i) the prefix hit does not introduce any additional error, since the label that the \gls{dl} model would predict $l_{S_K}$ is the same as the one that is stored in the cache for the prefix series $l_{S_\delta}$ (that we denote as {\it Good});  or,  (ii) the approximate hit introduces an additional error (that sums up to the DL model error) as the label that the \gls{dl} model would predict $l_{S_K}$ is not the same as the one that is stored in the cache for the prefix series $l_{S_\delta}$ (that we denote as {\it Error}).

To evaluate the performance of the caching system in realistic network scenarios we simulated a workload by using a dataset containing \SI{\nbflows}{\kilo\nothing} flows extracted from both \emph{campus} and \emph{access} ones for which we have corresponding labels \ie{} the ground truth used to train the \gls{dl} model. \Cref{fig:hit-ratio} presents the hit ratio for different caching strategies where regular cache means that the key used for the cache is $S_K$ with $K=10$ and prefix 4 and 6 means that the key used for caching analytics results is $S_\delta$ with $\delta = 4$ and $6$ respectively. The gain brought by the short prefix $\delta=4$ with respect to regular cache is significant with an increase of hit-ratio by $3\times$ although it starts affecting classification accuracy. More conservative settings such as $\delta=6$ allows a reduction of the \gls{dl} workload by approximately \SI{30}{\percent}, limiting the error rate to less than \SI{1}{\percent}. We further analyze the performance of caching in \Cref{sec:am_bench}.

\section{System evaluation}
\label{sec:am_bench}

\paragraph{Scenario} We evaluated the overall performance of \acrshort{fix} as a flow classifier that receives packets in the input port and outputs them in the opposite direction with a tag indicating the class (we use the IP field Type of Service) when the analytics is available. Our evaluation ran on two directly connected servers. The characteristics (cf. \Cref{sec:rqrmnt}) are: each server is equipped with Intel Xeon Platinum 8164 CPUs @ 2.00GHz (L1/L2/L3 caches 32 data+32 instruction)/1024/36608 kB), a \SI{100}{\giga bps} Mellanox MCX515A-CCAT ConnectX-5 NIC and a Huawei Atlas 300I:3010 Inference Card. In the test, one server ran MoonGen \cite{moongen} while the other one ran \acrshort{fix} with data array size and buckets equal to $2^{22}$ (\SI{4}{\mega\nothing}), and stale timeout \SI{30}{s}. Note that although servers running \acrshort{fix} are a high-end server we only use a limited amount of resources \eg{} 2/4 cores, 1/2 TPU chips, and a few MB of DRAM that can be found (or installed) at the network access. The workload is two-minute-long traces replayed by MoonGen. 
Since the throughput of our original traces is too low, we generated realistic traces as follows: the \emph{flow} arrival process follows a Poisson process with average $\lambda$, while the \textit{packets} conform to the real flow characteristics (\ie{} inter-packet delay, size, and the number of packets per flow) randomly chosen from a catalog of \SI{1}{\mega \nothing} flows extracted from the Access dataset. 

\paragraph{Dynamic batching} We started our evaluation by a scenario in which a single 1:1:1 \acrshort{fix} pipeline (\ie{} a flow manager, an ascend manager, and a \gls{tpu}) was given \SI{50}{\kilo flows/s}. \Cref{fig:carryon} compare timeout and \carry{} batching policies in terms of, from left to right, time to perform the analytics since \ith{10} packet 
, padding ratio
, and ratio of flows that did not get analytics before their end \ie{} post-mortem. 
The lower the timeout, the smaller the time to get the label. However, we point out that a smaller timeout drives the system to an inefficient operational point, especially for low arrival rate. On the contrary, a higher timeout leads to a higher percentage of post-mortem analytics. Finally, as pointed out in \Cref{sec:batching}, the \carry{} batching policy helps to control the level of padding trading off additional delay for the time to get the label in some cases \ie{} \ith{99} and \ith{75} percentiles. The lower the $\Phi$, the lower the padding and waste of processing and the higher the additional delay.

\paragraph{Approximate caching}  In the same scenario we also evaluated the impact of the caching module in \acrshort{fix} configured with timeout \SI{10}{\milli\second} and $\Phi=\num{0.2}$ in case of \carry{}. For this scenario, we dimensioned the approximate \gls{lru} cache with $\delta=6$ to have a hit ratio of \num{0.3} and we do not report here the difference between regular and approximate prefix cache as they would achieve similar results but with the advantage of a smaller memory footprint for the latter, at the price of a higher error. \Cref{fig:withcache} reports the results we obtained with timeout and \carry{} batching policies with or without caching in terms of time to get the label, padding, and post-mortem analytics. 
The results show that, for both batching strategies, the time to get the label decreases as some of the series \ie{} the one for which analytics is cached, are retrieved much faster. Notice also that, for the same reason the variability of the delay increases as well. At the same time, given the fact that the delay to get the label decreases, also the post-mortem analytics ratio decreases.

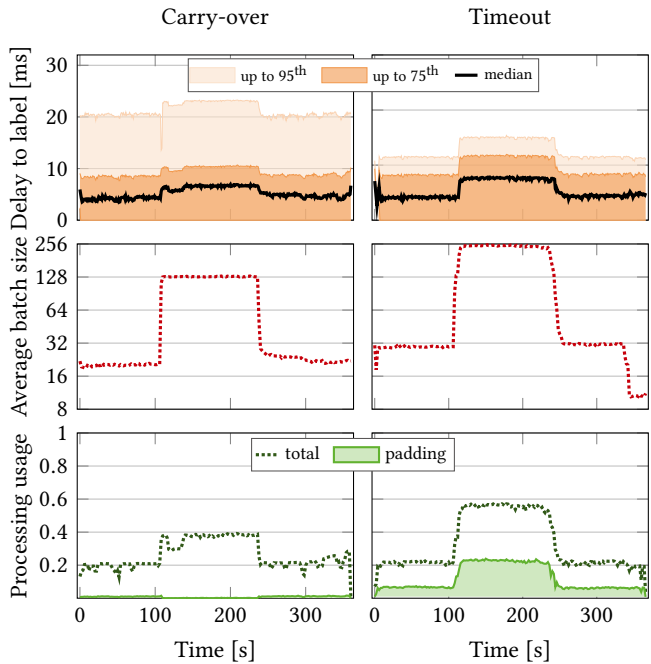
\begin{figure}
    \centering
    \begin{tikzpicture}

\pgfplotstableread[col sep = comma]
    {Data/dynamic-test-70k/df_batchsize-carryon.csv}
    \tabBatchCarryon
    
\pgfplotstableread[col sep = comma]
    {Data/dynamic-test-70k/df_batchsize-fixed.csv}
    \tabBatchFixed
    
\pgfplotstableread[col sep = comma]
    {Data/dynamic-test-70k/df_processing-carryon.csv}
    \tabProcessingCarryon

\pgfplotstableread[col sep = comma]
    {Data/dynamic-test-70k/df_processing-fixed.csv}
    \tabProcessingFixed

\pgfplotstableread[col sep = comma]
    {Data/dynamic-test-70k/df_ttlabel-carryon.csv}
    \tabTtlabelCarryon

\pgfplotstableread[col sep = comma]
    {Data/dynamic-test-70k/df_ttlabel-fixed.csv}
    \tabTtlabelFixed

\def\totalproc{0.00000001}
\def\ymax{1}

\begin{groupplot}[
    group style={
        group name=my plots,
        group size=2 by 3,
        x descriptions at=edge bottom,
        y descriptions at=edge left,
        horizontal sep=7pt,
        vertical sep=9pt,
    },
    /tikz/font=\small,
    width=0.62*\columnwidth,
    height=1.35*\heightfigure,
    xlabel = {Time [s]},
    ymajorgrids,
    ylabel near ticks,
    enlarge y limits = 0,
    enlarge x limits = 0.01,
    ylabel style = {
        at={(axis description cs:-0.13,.5)}
    },
    ]
    

\nextgroupplot[
    title= \carry,
    ymin = 0,
    ymax = 32,
    ylabel = {Delay to label [ms]\vphantom{g}},
    legend pos=north east,
    legend style={
        font=\tiny,
        anchor = north west,
        at = {(0.4,0.98)},
        /tikz/every even column/.append style={
            column sep=3pt},
        fill=white,
        inner sep=0.4pt,
        draw=colorGray},
    legend columns = 3,
    ]
    
\addplot[
    name path=mymax,
    very thin,
    no markers,
    colorOrange!30,
    ] table [x=x,
        y expr= 0.001*\thisrow{q95}
        ]   {\tabTtlabelCarryon};
        
\addplot[
    name path=mymin,
    very thin,
    no markers,
    colorOrange!30,
    ] table [x=x,
        y expr= 0.001*\thisrow{q05}
        ]   {\tabTtlabelCarryon};

  \addplot [colorOrange!20,
    fill opacity=0.6] fill between[ 
    of = mymin and mymax];
    
\addplot[
    name path=myuw,
    very thin,
    no markers,
    colorOrange!70,
    ] table [x=x,
        y expr= 0.001*\thisrow{q75}
        ]   {\tabTtlabelCarryon};
        
\addplot[
    name path=mylw,
    very thin,
    no markers,
    colorOrange!70,
    ] table [x=x,
        y expr= 0.001*\thisrow{q25}
        ]   {\tabTtlabelCarryon};

  \addplot [colorOrange!70,
  fill opacity=0.6] fill between[ 
    of = mylw and myuw];


\addplot[very thick, 
        draw=black,
        ] 
    table[
        x=x,
        y expr= 0.001*\thisrow{q50}
    ]
    {\tabTtlabelCarryon};
\legend{
    ,
    ,
    up to \ith{95}, 
    ,
    ,
    up to \ith{75}, 
    median
}

\nextgroupplot[
    title = Timeout\vphantom{y},
    ymin = 0,
    ymax = 30]
    
\addplot[
    name path=mymax,
    very thin,
    no markers,
    colorOrange!30,
    ] table [x=x,
        y expr= 0.001*\thisrow{q95}
        ]  {\tabTtlabelFixed};
        
\addplot[
    name path=mymin,
    very thin,
    no markers,
    colorOrange!30,
    ] table [x=x,
        y expr= 0.001*\thisrow{q05}
        ]  {\tabTtlabelFixed};

  \addplot [colorOrange!20,
    fill opacity=0.6] fill between[ 
    of = mymin and mymax];
    
\addplot[
    name path=myuw,
    very thin,
    no markers,
    colorOrange!70,
    ] table [x=x,
        y expr= 0.001*\thisrow{q75}
        ]  {\tabTtlabelFixed};
        
\addplot[
    name path=mylw,
    very thin,
    no markers,
    colorOrange!70,
    ] table [x=x,
        y expr= 0.001*\thisrow{q25}
        ]   {\tabTtlabelFixed};

  \addplot [colorOrange!70,
  fill opacity=0.6] fill between[ 
    of = mylw and myuw];


\addplot[very thick, 
        draw=black,
        ] 
    table[
        x=x,
        y expr= 0.001*\thisrow{q50}
    ]
    {\tabTtlabelFixed};

\nextgroupplot[
    ymode=log, 
    log ticks with fixed point,
    scaled y ticks=false,
    ymin=8,
    ymax=256,
    ytick = {8, 16, 32, 64, 128, 256},
    ylabel = Average batch size]
    
\addplot[very thick, 
        draw=colorRed,
        densely dotted] table[
        x=x,
        y=batchsize]
    {\tabBatchCarryon};

\nextgroupplot[
    ymode=log, 
    log ticks with fixed point,
    scaled y ticks=false,
    ymin=8,
    ymax=256,
    ytick = {8, 16, 32, 64, 128, 256},
    ]
    
\addplot[very thick, 
        draw=colorRed,
        densely dotted] table[
        x=x,
        y=batchsize]
    {\tabBatchFixed};
    

\nextgroupplot[
    ymin = 0,
    ymax = \ymax,
    ytick={0.2,0.4,0.6, 0.8, 1},
    ylabel = Processing usage,
    ]
    
\addplot[very thick, 
        draw=colorGreen!50!black,
        densely dotted] 
    table[
        x=x,
        y expr=\totalproc*\thisrow{ttlabel}       
    ]
    {\tabProcessingCarryon};
    
\addplot[thick, 
        draw=colorGreen,
        fill=colorGreen!30,
        area legend] 
    table[
        x=x,
        y expr=\totalproc*\thisrow{padding}       
    ]
    {\tabProcessingCarryon};

\nextgroupplot[
    ymin = 0,
    ytick={0.2,0.4,0.6, 0.8, 1},
    ymax = \ymax,
    legend style={
        font=\scriptsize,
        anchor = north east,
        at = {(0.3,0.97)},
        /tikz/every even column/.append style={
            column sep=3pt},
        fill=white,
        inner sep=1pt,
        draw=colorGray},
    legend columns = 2]
    
\addplot[very thick, 
        draw=colorGreen!50!black,
        densely dotted] 
    table[
        x=x,
        y expr=\totalproc*\thisrow{ttlabel}       
    ]
    {\tabProcessingFixed};
\addplot[thick, 
        draw=colorGreen,
        fill=colorGreen!30,
        area legend] 
    table[
        x=x,
        y expr=\totalproc*\thisrow{padding}       
    ]
    {\tabProcessingFixed};
\legend{total, padding}

\end{groupplot}

\end{tikzpicture}
    \caption{Fixed and \carry{} batching strategies over time under dynamic traffic conditions.}
    \label{fig:dynamic}
\end{figure}
\paragraph{Flash crowd} Finally we evaluated \acrshort{fix} performance in a dynamic flash-crowd scenario in which two 1:1:1 pipelines are subject to an input traffic at \SI{10}{\kilo flows/s} for the first \num{120} seconds, then \SI{70}{\kilo flows/s} in the next \num{120} seconds, and \SI{10}{\kilo flows/s} for the last part of the experiment. \acrshort{fix} was configured with a prefix $\delta=6$ LRU cache and size for which the hit ratio is approximately \num{0.3}. The \carry{} strategy is set with  \SI{10}{ms} timeout and $\Phi=\num{0.2}$. 
\Cref{fig:dynamic} presents the time to get the label, the ratio of \gls{tpu} usage, and the average batch size over time. Both timeout and \carry{} batching strategies are able to scale up batch size and consequently the processing usage when the traffic load suddenly increases. Notice that, despite the higher processing, the time to get the label only slightly increases. Finally, we highlight the fact that \carry{} is successful in dynamically adapting to the right batch size, reduces energy consumption by lower processing power, and further avoid wasting processing power by greatly limiting the processing power spent for padded input analysis.


\section{Related Work}
\label{sec:related}
The success of \gls{dl} technology has ignited interest for its in-network  use,
so that valuable work started tackling the issue  of \gls{dl} analytics offloading
from a networking perspective. 

\textit{Server-side offloading:}
The \gls{ml} community has extensively studied the performance of hardware accelerators~\cite{HPECsurvey19,cogmi19} and the design of offloading mechanisms for data inference to external hardware. However, the existing \emph{inference serving} systems~\cite{amazonei} and literature~\cite{clipper,WangWGZCNOS18,ZhangYWY19,LeeSCSWI18} generally target the case of offloading image recognition to GPU-equipped servers in a datacenter. The closest work in this space is \emph{Clipper}~\cite{clipper}, whose design also leverages caching and batching, as rather typical weapons in the design space. This solution and subsequent work~\cite{WangWGZCNOS18} target cloud-based inference with desirable latency targets (around \SI{20}{ms}). Recent work \cite{nigade2020clownfish} also studies the cooperation among edge and cloud in the context of video streaming analytics. However, as pointed out in~\cite{XiongZ19}, the volume of data and processing time for \gls{dl} inference models on network traffic is radically different from that of computer vision applications, so that the system requirement, the processed inputs, and the usage of the analytics output are radically different as well.

\textit{Network-wide offloading:}
Complementary to our work, programmable network devices have been used to assist the processing of heavy workloads~\cite{MiaoZKLY17,SanvitoSB18,SapioAACK17} some of which have a specific focus on network-assisted DL offloading~\cite{SanvitoSB18}.  For instance, BananaSplit~\cite{SanvitoSB18} focuses on breaking down DL computation of complex image recognition models across multiple network devices using SmartNIC whereas in our work we adopt the opposite viewpoint and target in-device execution of DL models applied to data plane traffic.

\textit{In-device machine learning (ML) offloading:}
Work that focuses on the programmable data-plane for switches based on \gls{pisa}~\cite{innetworkNN18,GonzalezMGMHNN17,SanvitoSB18,XiongZ19} tries to cope with the limited resources that are available for the implementation of \gls{ml} models.
For example \cite{XiongZ19} adapt several \gls{ml} models to the match-action table model in P4, while \cite{abs-1909-05680} implementing a random forest model to classify traffic.
In both cases, the benefits of running the analytics in the data-plane are limited since no immediate packet forwarding decision derives from the analytics. Moreover, the lack of computing and memory resources prevents the implementation of \gls{dl} models altogether. 

\textit{In-device  deep learning (DL) offloading:}
Offloading \gls{dl} inference to hardware accelerators brings better accuracy, the ability to run several \gls{dl} models in parallel, and to leverage specialized \gls{dl} frameworks such as TensorFlow~\cite{AbadiBCCDDDGIIK16}.
Closest to our work are thus two recent (not peer-reviewed) papers, which explore in-device \gls{dl} offloading. ASIC is used in \cite{swamy2020taurus} for \gls{dl} inference at packet level but only on toy-models with \num{3} layers and \num{21} neurons, \ie{} $5000\times$ smaller than the model we use. Smart \gls{nic} is used in~\cite{abs-2009-02353}, that however limits model size to \num{50} \textit{binary} neurons, \ie{} $2000\times$ fewer weights, each with a resolution $32\times$ smaller than in our case study. To attain sub-microsecond latency, \cite{swamy2020taurus,abs-2009-02353} restrict themselves to such tiny models that it becomes questionable if their execution can have any practical use given the significant distance of such shallow models from the depth needed to embrace the expected benefits of \gls{dl}. Our work takes the opposite viewpoint and tackles a timely and efficient execution of relevant \gls{dl} models for edge intelligence by using the appropriate offloading hardware, \ie{} \glspl{tpu}.

\section{Conclusion}
\acrshort{fix} is the first system to integrate forwarding and advanced analytics capabilities, exploiting \gls{tpu} hardware acceleration to offer efficient execution of Deep Learning analytics in network devices' data path. Its system design leverages asynchronous communication between forwarding and analytics engines, optimizing the usage of the hardware accelerator by introducing novel dynamic batching and smart caching policies. This makes \acrshort{fix} capable of high-speed (100Gbps), low-delay (below 10ms), and low-power consumption (on the order of few tens of Watts for the \gls{tpu}) on off-the-shelf hardware, ultimately paving the way to the deployment of embedded intelligence at the network edge.


{


\bibliographystyle{ACM-Reference-Format}
\bibliography{reference}}


\begin{thebibliography}{47}


\ifx \showCODEN    \undefined \def \showCODEN     #1{\unskip}     \fi
\ifx \showDOI      \undefined \def \showDOI       #1{#1}\fi
\ifx \showISBNx    \undefined \def \showISBNx     #1{\unskip}     \fi
\ifx \showISBNxiii \undefined \def \showISBNxiii  #1{\unskip}     \fi
\ifx \showISSN     \undefined \def \showISSN      #1{\unskip}     \fi
\ifx \showLCCN     \undefined \def \showLCCN      #1{\unskip}     \fi
\ifx \shownote     \undefined \def \shownote      #1{#1}          \fi
\ifx \showarticletitle \undefined \def \showarticletitle #1{#1}   \fi
\ifx \showURL      \undefined \def \showURL       {\relax}        \fi
\providecommand\bibfield[2]{#2}
\providecommand\bibinfo[2]{#2}
\providecommand\natexlab[1]{#1}
\providecommand\showeprint[2][]{arXiv:#2}

\bibitem[\protect\citeauthoryear{??}{hua}{2018}]%
        {huaweinic}
 \bibinfo{year}{2018}\natexlab{}.
\newblock \bibinfo{title}{{Huawei IN Series Intelligent NICs}}.
\newblock
  \bibinfo{howpublished}{\url{https://e.huawei.com/en/products/servers/pcie-ssd/in-card}}.
\newblock


\bibitem[\protect\citeauthoryear{??}{ama}{2020}]%
        {amazonei}
 \bibinfo{year}{2020}\natexlab{}.
\newblock \bibinfo{title}{Amazon Elastic Inference}.
\newblock
  \bibinfo{howpublished}{\url{https://docs.aws.amazon.com/elastic-inference/}}.
\newblock


\bibitem[\protect\citeauthoryear{??}{asc}{2020}]%
        {ascend310}
 \bibinfo{year}{2020}\natexlab{}.
\newblock \bibinfo{title}{{Ascend 310 chip}}.
\newblock
  \bibinfo{howpublished}{\url{https://e.huawei.com/se/products/cloud-computing-dc/atlas/ascend-310}}.
\newblock


\bibitem[\protect\citeauthoryear{??}{cor}{2020}]%
        {coral}
 \bibinfo{year}{2020}\natexlab{}.
\newblock \bibinfo{title}{{Google Coral}}.
\newblock \bibinfo{howpublished}{\url{https://coral.ai/products/}}.
\newblock


\bibitem[\protect\citeauthoryear{??}{min}{2020}]%
        {mindstudio}
 \bibinfo{year}{2020}\natexlab{}.
\newblock \bibinfo{title}{Huawei MindStudio}.
\newblock
  \bibinfo{howpublished}{\url{https://www.huaweicloud.com/intl/en-us/ascend/doc/mindstudio}}.
\newblock


\bibitem[\protect\citeauthoryear{??}{mel}{2020a}]%
        {mellanoxnic}
 \bibinfo{year}{2020}\natexlab{a}.
\newblock \bibinfo{title}{{Mellanox BlueField}}.
\newblock
  \bibinfo{howpublished}{\url{https://www.mellanox.com/products/smartnic}}.
\newblock


\bibitem[\protect\citeauthoryear{??}{mel}{2020b}]%
        {mellanox}
 \bibinfo{year}{2020}\natexlab{b}.
\newblock \bibinfo{title}{{Nividia EGX A100}}.
\newblock
  \bibinfo{howpublished}{\url{https://www.nvidia.com/en-us/data-center/products/egx-a100/}}.
\newblock


\bibitem[\protect\citeauthoryear{??}{tri}{2020}]%
        {triton}
 \bibinfo{year}{2020}\natexlab{}.
\newblock \bibinfo{title}{NVidia Triton Inference Server}.
\newblock
  \bibinfo{howpublished}{\url{https://docs.nvidia.com/deeplearning/triton-inference-server/user-guide/}}.
\newblock


\bibitem[\protect\citeauthoryear{Abadi, Barham, Chen, Chen, Davis, Dean, Devin,
  Ghemawat, Irving, Isard, Kudlur, Levenberg, Monga, Moore, Murray, Steiner,
  Tucker, Vasudevan, Warden, Wicke, Yu, and Zheng}{Abadi et~al\mbox{.}}{2016}]%
        {AbadiBCCDDDGIIK16}
\bibfield{author}{\bibinfo{person}{Mart{\'{\i}}n Abadi}, \bibinfo{person}{Paul
  Barham}, \bibinfo{person}{Jianmin Chen}, \bibinfo{person}{Zhifeng Chen},
  \bibinfo{person}{Andy Davis}, \bibinfo{person}{Jeffrey Dean},
  \bibinfo{person}{Matthieu Devin}, \bibinfo{person}{Sanjay Ghemawat},
  \bibinfo{person}{Geoffrey Irving}, \bibinfo{person}{Michael Isard},
  \bibinfo{person}{Manjunath Kudlur}, \bibinfo{person}{Josh Levenberg},
  \bibinfo{person}{Rajat Monga}, \bibinfo{person}{Sherry Moore},
  \bibinfo{person}{Derek~Gordon Murray}, \bibinfo{person}{Benoit Steiner},
  \bibinfo{person}{Paul~A. Tucker}, \bibinfo{person}{Vijay Vasudevan},
  \bibinfo{person}{Pete Warden}, \bibinfo{person}{Martin Wicke},
  \bibinfo{person}{Yuan Yu}, {and} \bibinfo{person}{Xiaoqiang Zheng}.}
  \bibinfo{year}{2016}\natexlab{}.
\newblock \showarticletitle{TensorFlow: {A} System for Large-Scale Machine
  Learning}. In \bibinfo{booktitle}{\emph{{USENIX} Symp. on Op. Sys. Design and
  Impl. {OSDI}}}. \bibinfo{pages}{265--283}.
\newblock


\bibitem[\protect\citeauthoryear{Aceto, Ciuonzo, Montieri, and
  Pescap{\`{e}}}{Aceto et~al\mbox{.}}{2019}]%
        {AcetoCMP19}
\bibfield{author}{\bibinfo{person}{Giuseppe Aceto}, \bibinfo{person}{Domenico
  Ciuonzo}, \bibinfo{person}{Antonio Montieri}, {and} \bibinfo{person}{Antonio
  Pescap{\`{e}}}.} \bibinfo{year}{2019}\natexlab{}.
\newblock \showarticletitle{Mobile Encrypted Traffic Classification Using Deep
  Learning: Experimental Evaluation, Lessons Learned, and Challenges}.
\newblock \bibinfo{journal}{\emph{{IEEE} Trans. Network and Service
  Management}} \bibinfo{volume}{16}, \bibinfo{number}{2}
  (\bibinfo{year}{2019}), \bibinfo{pages}{445--458}.
\newblock


\bibitem[\protect\citeauthoryear{Albanie}{Albanie}{2019}]%
        {convnet}
\bibfield{author}{\bibinfo{person}{Samuel Albanie}.}
  \bibinfo{year}{2019}\natexlab{}.
\newblock \bibinfo{title}{Memory consumption and FLOP count estimates for
  Convolutional Neural Networks}.
\newblock
  \bibinfo{howpublished}{\url{https://github.com/albanie/convnet-burden}}.
\newblock


\bibitem[\protect\citeauthoryear{Beliard, Finamore, and Rossi}{Beliard
  et~al\mbox{.}}{2020}]%
        {pandorabox}
\bibfield{author}{\bibinfo{person}{Cedric Beliard}, \bibinfo{person}{Alessandro
  Finamore}, {and} \bibinfo{person}{Dario Rossi}.}
  \bibinfo{year}{2020}\natexlab{}.
\newblock \bibinfo{title}{Opening the Deep Pandora Box: Explainable Traffic
  Classification}.
\newblock
\newblock


\bibitem[\protect\citeauthoryear{Bernaille, Teixeira, Akodkenou, Soule, and
  Salamatian}{Bernaille et~al\mbox{.}}{2006}]%
        {teixeira06ccr}
\bibfield{author}{\bibinfo{person}{Laurent Bernaille}, \bibinfo{person}{Renata
  Teixeira}, \bibinfo{person}{Ismael Akodkenou}, \bibinfo{person}{Augustin
  Soule}, {and} \bibinfo{person}{Kav{\'{e}} Salamatian}.}
  \bibinfo{year}{2006}\natexlab{}.
\newblock \showarticletitle{Traffic Classification on the Fly}.
\newblock \bibinfo{journal}{\emph{ACM SIGCOMM Comput. Commun. Rev.}}
  \bibinfo{volume}{36}, \bibinfo{number}{2} (\bibinfo{year}{2006}),
  \bibinfo{pages}{23–26}.
\newblock


\bibitem[\protect\citeauthoryear{Bin, David, and Michael}{Bin
  et~al\mbox{.}}{2013}]%
        {memc3}
\bibfield{author}{\bibinfo{person}{Fan Bin}, \bibinfo{person}{G.~Andersen
  David}, {and} \bibinfo{person}{Kaminsky Michael}.}
  \bibinfo{year}{2013}\natexlab{}.
\newblock \showarticletitle{MemC3: Compact and Concurrent MemCache with Dumber
  Caching and Smarter Hashing}. In \bibinfo{booktitle}{\emph{Proc. of {USENIX}
  {NSDI}}}. \bibinfo{pages}{371--384}.
\newblock


\bibitem[\protect\citeauthoryear{Busse{-}Grawitz, Meier, Dietm{\"{u}}ller,
  B{\"{u}}hler, and Vanbever}{Busse{-}Grawitz et~al\mbox{.}}{2019}]%
        {abs-1909-05680}
\bibfield{author}{\bibinfo{person}{Coralie Busse{-}Grawitz},
  \bibinfo{person}{Roland Meier}, \bibinfo{person}{Alexander Dietm{\"{u}}ller},
  \bibinfo{person}{Tobias B{\"{u}}hler}, {and} \bibinfo{person}{Laurent
  Vanbever}.} \bibinfo{year}{2019}\natexlab{}.
\newblock \showarticletitle{pForest: In-Network Inference with Random Forests}.
\newblock \bibinfo{journal}{\emph{CoRR}} (\bibinfo{year}{2019}).
\newblock
\showeprint[arxiv]{1909.05680}
\urldef\tempurl%
\url{http://arxiv.org/abs/1909.05680}
\showURL{%
\tempurl}


\bibitem[\protect\citeauthoryear{Crankshaw, Wang, Zhou, Franklin, Gonzalez, and
  Stoica}{Crankshaw et~al\mbox{.}}{2017}]%
        {clipper}
\bibfield{author}{\bibinfo{person}{Daniel Crankshaw}, \bibinfo{person}{Xin
  Wang}, \bibinfo{person}{Giulio Zhou}, \bibinfo{person}{Michael~J. Franklin},
  \bibinfo{person}{Joseph~E. Gonzalez}, {and} \bibinfo{person}{Ion Stoica}.}
  \bibinfo{year}{2017}\natexlab{}.
\newblock \showarticletitle{Clipper: A Low-Latency Online Prediction Serving
  System}. In \bibinfo{booktitle}{\emph{USENIX NSDI}}.
\newblock


\bibitem[\protect\citeauthoryear{Crotti, Dusi, Gringoli, and Salgarelli}{Crotti
  et~al\mbox{.}}{2007}]%
        {crotti07ccr}
\bibfield{author}{\bibinfo{person}{Manuel Crotti}, \bibinfo{person}{Maurizio
  Dusi}, \bibinfo{person}{Francesco Gringoli}, {and} \bibinfo{person}{Luca
  Salgarelli}.} \bibinfo{year}{2007}\natexlab{}.
\newblock \showarticletitle{Traffic classification through simple statistical
  fingerprinting}.
\newblock \bibinfo{journal}{\emph{ACM SIGCOMM Comput. Commun. Rev.}}
  \bibinfo{volume}{37}, \bibinfo{number}{1} (\bibinfo{year}{2007}),
  \bibinfo{pages}{5--16}.
\newblock


\bibitem[\protect\citeauthoryear{{Dai} and {Berleant}}{{Dai} and
  {Berleant}}{2019}]%
        {cogmi19}
\bibfield{author}{\bibinfo{person}{W. {Dai}} {and} \bibinfo{person}{D.
  {Berleant}}.} \bibinfo{year}{2019}\natexlab{}.
\newblock \showarticletitle{Benchmarking Contemporary Deep Learning Hardware
  and Frameworks: A Survey of Qualitative Metrics}. In
  \bibinfo{booktitle}{\emph{2019 IEEE First International Conference on
  Cognitive Machine Intelligence (CogMI)}}. \bibinfo{pages}{148--155}.
\newblock


\bibitem[\protect\citeauthoryear{David and Netanyahu}{David and
  Netanyahu}{2015}]%
        {DavidN15}
\bibfield{author}{\bibinfo{person}{Omid~E. David} {and}
  \bibinfo{person}{Nathan~S. Netanyahu}.} \bibinfo{year}{2015}\natexlab{}.
\newblock \showarticletitle{DeepSign: Deep learning for automatic malware
  signature generation and classification}. In \bibinfo{booktitle}{\emph{Int.
  Joint Conf. on Neural Networks ({IJCNN})}}. \bibinfo{pages}{1--8}.
\newblock


\bibitem[\protect\citeauthoryear{{Emmerich}, {Gallenm{\"u}ller}, {Raumer},
  {Wohlfart}, and {Carle}}{{Emmerich} et~al\mbox{.}}{2015}]%
        {moongen}
\bibfield{author}{\bibinfo{person}{Paul {Emmerich}}, \bibinfo{person}{Sebastian
  {Gallenm{\"u}ller}}, \bibinfo{person}{Daniel {Raumer}},
  \bibinfo{person}{Florian {Wohlfart}}, {and} \bibinfo{person}{Georg {Carle}}.}
  \bibinfo{year}{2015}\natexlab{}.
\newblock \showarticletitle{{MoonGen: A Scriptable High-Speed Packet
  Generator}}. In \bibinfo{booktitle}{\emph{ACM Internet Measurement Conf.
  ({IMC})}}.
\newblock


\bibitem[\protect\citeauthoryear{Falchi, Lucchese, Orlando, Perego, and
  Rabitti}{Falchi et~al\mbox{.}}{2012}]%
        {FalchiLOPR12}
\bibfield{author}{\bibinfo{person}{Fabrizio Falchi}, \bibinfo{person}{Claudio
  Lucchese}, \bibinfo{person}{Salvatore Orlando}, \bibinfo{person}{Raffaele
  Perego}, {and} \bibinfo{person}{Fausto Rabitti}.}
  \bibinfo{year}{2012}\natexlab{}.
\newblock \showarticletitle{Similarity caching in large-scale image retrieval}.
\newblock \bibinfo{journal}{\emph{Inf. Process. Manag.}} \bibinfo{volume}{48},
  \bibinfo{number}{5} (\bibinfo{year}{2012}), \bibinfo{pages}{803--818}.
\newblock


\bibitem[\protect\citeauthoryear{Gonzalez, Manco, Garc{\'{\i}}a{-}Dur{\'{a}}n,
  Mendes, Huici, Niccolini, and Niepert}{Gonzalez et~al\mbox{.}}{2017}]%
        {GonzalezMGMHNN17}
\bibfield{author}{\bibinfo{person}{Roberto Gonzalez}, \bibinfo{person}{Filipe
  Manco}, \bibinfo{person}{Alberto Garc{\'{\i}}a{-}Dur{\'{a}}n},
  \bibinfo{person}{Jose Mendes}, \bibinfo{person}{Felipe Huici},
  \bibinfo{person}{Saverio Niccolini}, {and} \bibinfo{person}{Mathias
  Niepert}.} \bibinfo{year}{2017}\natexlab{}.
\newblock \showarticletitle{Net2Vec: Deep Learning for the Network}. In
  \bibinfo{booktitle}{\emph{Proc. of Big-DAMA Workshop at ACM Sigcomm}}.
\newblock


\bibitem[\protect\citeauthoryear{Jin, Guo, Xiao, Shi, Niu, Liu, Qian, and
  Wang}{Jin et~al\mbox{.}}{2019}]%
        {JinGXSNLQW19}
\bibfield{author}{\bibinfo{person}{Panpan Jin}, \bibinfo{person}{Jian Guo},
  \bibinfo{person}{Yikai Xiao}, \bibinfo{person}{Rong Shi},
  \bibinfo{person}{Yipei Niu}, \bibinfo{person}{Fangming Liu},
  \bibinfo{person}{Chen Qian}, {and} \bibinfo{person}{Yang Wang}.}
  \bibinfo{year}{2019}\natexlab{}.
\newblock \showarticletitle{PostMan: Rapidly Mitigating Bursty Traffic by
  Offloading Packet Processing}. In \bibinfo{booktitle}{\emph{{USENIX} {ATC}}}.
  \bibinfo{pages}{849--862}.
\newblock


\bibitem[\protect\citeauthoryear{Jouppi, Young, Patil, Patterson, Agrawal,
  Bajwa, Bates, Bhatia, Boden, Borchers, Boyle, Cantin, Chao, Clark, Coriell,
  Daley, Dau, Dean, Gelb, Ghaemmaghami, Gottipati, Gulland, Hagmann, Ho,
  Hogberg, Hu, Hundt, Hurt, Ibarz, Jaffey, Jaworski, Kaplan, Khaitan,
  Killebrew, Koch, Kumar, Lacy, Laudon, Law, Le, Leary, Liu, Lucke, Lundin,
  MacKean, Maggiore, Mahony, Miller, Nagarajan, Narayanaswami, Ni, Nix, Norrie,
  Omernick, Penukonda, Phelps, Ross, Ross, Salek, Samadiani, Severn, Sizikov,
  Snelham, Souter, Steinberg, Swing, Tan, Thorson, Tian, Toma, Tuttle,
  Vasudevan, Walter, Wang, Wilcox, and Yoon}{Jouppi et~al\mbox{.}}{2017}]%
        {JouppiYPPABBBBB17}
\bibfield{author}{\bibinfo{person}{Norman~P. Jouppi}, \bibinfo{person}{Cliff
  Young}, \bibinfo{person}{Nishant Patil}, \bibinfo{person}{David~A.
  Patterson}, \bibinfo{person}{Gaurav Agrawal}, \bibinfo{person}{Raminder
  Bajwa}, \bibinfo{person}{Sarah Bates}, \bibinfo{person}{Suresh Bhatia},
  \bibinfo{person}{Nan Boden}, \bibinfo{person}{Al Borchers},
  \bibinfo{person}{Rick Boyle}, \bibinfo{person}{Pierre{-}luc Cantin},
  \bibinfo{person}{Clifford Chao}, \bibinfo{person}{Chris Clark},
  \bibinfo{person}{Jeremy Coriell}, \bibinfo{person}{Mike Daley},
  \bibinfo{person}{Matt Dau}, \bibinfo{person}{Jeffrey Dean},
  \bibinfo{person}{Ben Gelb}, \bibinfo{person}{Tara~Vazir Ghaemmaghami},
  \bibinfo{person}{Rajendra Gottipati}, \bibinfo{person}{William Gulland},
  \bibinfo{person}{Robert Hagmann}, \bibinfo{person}{C.~Richard Ho},
  \bibinfo{person}{Doug Hogberg}, \bibinfo{person}{John Hu},
  \bibinfo{person}{Robert Hundt}, \bibinfo{person}{Dan Hurt},
  \bibinfo{person}{Julian Ibarz}, \bibinfo{person}{Aaron Jaffey},
  \bibinfo{person}{Alek Jaworski}, \bibinfo{person}{Alexander Kaplan},
  \bibinfo{person}{Harshit Khaitan}, \bibinfo{person}{Daniel Killebrew},
  \bibinfo{person}{Andy Koch}, \bibinfo{person}{Naveen Kumar},
  \bibinfo{person}{Steve Lacy}, \bibinfo{person}{James Laudon},
  \bibinfo{person}{James Law}, \bibinfo{person}{Diemthu Le},
  \bibinfo{person}{Chris Leary}, \bibinfo{person}{Zhuyuan Liu},
  \bibinfo{person}{Kyle Lucke}, \bibinfo{person}{Alan Lundin},
  \bibinfo{person}{Gordon MacKean}, \bibinfo{person}{Adriana Maggiore},
  \bibinfo{person}{Maire Mahony}, \bibinfo{person}{Kieran Miller},
  \bibinfo{person}{Rahul Nagarajan}, \bibinfo{person}{Ravi Narayanaswami},
  \bibinfo{person}{Ray Ni}, \bibinfo{person}{Kathy Nix},
  \bibinfo{person}{Thomas Norrie}, \bibinfo{person}{Mark Omernick},
  \bibinfo{person}{Narayana Penukonda}, \bibinfo{person}{Andy Phelps},
  \bibinfo{person}{Jonathan Ross}, \bibinfo{person}{Matt Ross},
  \bibinfo{person}{Amir Salek}, \bibinfo{person}{Emad Samadiani},
  \bibinfo{person}{Chris Severn}, \bibinfo{person}{Gregory Sizikov},
  \bibinfo{person}{Matthew Snelham}, \bibinfo{person}{Jed Souter},
  \bibinfo{person}{Dan Steinberg}, \bibinfo{person}{Andy Swing},
  \bibinfo{person}{Mercedes Tan}, \bibinfo{person}{Gregory Thorson},
  \bibinfo{person}{Bo Tian}, \bibinfo{person}{Horia Toma},
  \bibinfo{person}{Erick Tuttle}, \bibinfo{person}{Vijay Vasudevan},
  \bibinfo{person}{Richard Walter}, \bibinfo{person}{Walter Wang},
  \bibinfo{person}{Eric Wilcox}, {and} \bibinfo{person}{Doe~Hyun Yoon}.}
  \bibinfo{year}{2017}\natexlab{}.
\newblock \showarticletitle{In-Datacenter Performance Analysis of a Tensor
  Processing Unit}. In \bibinfo{booktitle}{\emph{Proc. of the 44th Symp. on
  Computer Architecture, {ISCA}}}. \bibinfo{pages}{1--12}.
\newblock


\bibitem[\protect\citeauthoryear{Lee, Scolari, Chun, Santambrogio, Weimer, and
  Interlandi}{Lee et~al\mbox{.}}{2018}]%
        {LeeSCSWI18}
\bibfield{author}{\bibinfo{person}{Yunseong Lee}, \bibinfo{person}{Alberto
  Scolari}, \bibinfo{person}{Byung{-}Gon Chun}, \bibinfo{person}{Marco~Domenico
  Santambrogio}, \bibinfo{person}{Markus Weimer}, {and} \bibinfo{person}{Matteo
  Interlandi}.} \bibinfo{year}{2018}\natexlab{}.
\newblock \showarticletitle{{PRETZEL:} Opening the Black Box of Machine
  Learning Prediction Serving Systems}. In \bibinfo{booktitle}{\emph{{USENIX}
  {OSDI}}}. \bibinfo{pages}{611--626}.
\newblock


\bibitem[\protect\citeauthoryear{Li, Niaki, Choffnes, Gill, and Mislove}{Li
  et~al\mbox{.}}{2019}]%
        {LiNCGM19}
\bibfield{author}{\bibinfo{person}{Fangfan Li}, \bibinfo{person}{Arian~Akhavan
  Niaki}, \bibinfo{person}{David~R. Choffnes}, \bibinfo{person}{Phillipa Gill},
  {and} \bibinfo{person}{Alan Mislove}.} \bibinfo{year}{2019}\natexlab{}.
\newblock \showarticletitle{A large-scale analysis of deployed traffic
  differentiation practices}. In \bibinfo{booktitle}{\emph{Proc. of {ACM}
  {SIGCOMM}}}. \bibinfo{pages}{130--144}.
\newblock


\bibitem[\protect\citeauthoryear{Mar{\'{\i}}n, Casas, and
  Capdehourat}{Mar{\'{\i}}n et~al\mbox{.}}{2018}]%
        {MarinCC18}
\bibfield{author}{\bibinfo{person}{Gonzalo Mar{\'{\i}}n},
  \bibinfo{person}{Pedro Casas}, {and} \bibinfo{person}{Germ{\'{a}}n
  Capdehourat}.} \bibinfo{year}{2018}\natexlab{}.
\newblock \showarticletitle{DeepSec meets RawPower - Deep Learning for
  Detection of Network Attacks Using Raw Representations}.
\newblock \bibinfo{journal}{\emph{{SIGMETRICS} Perform. Evaluation Rev.}}
  \bibinfo{volume}{46}, \bibinfo{number}{3} (\bibinfo{year}{2018}),
  \bibinfo{pages}{147--150}.
\newblock


\bibitem[\protect\citeauthoryear{Mestres, Rodr{\'{\i}}guez{-}Natal, Carner,
  Barlet{-}Ros, Alarc{\'{o}}n, Sol{\'{e}}, Munt{\'{e}}s{-}Mulero, Meyer,
  Barkai, Hibbett, Estrada, Maruf, Coras, Ermagan, Latapie, Cassar, Evans,
  Maino, Walrand, and Cabellos}{Mestres et~al\mbox{.}}{2017}]%
        {MestresRCBASMMB17}
\bibfield{author}{\bibinfo{person}{Albert Mestres}, \bibinfo{person}{Alberto
  Rodr{\'{\i}}guez{-}Natal}, \bibinfo{person}{Josep Carner},
  \bibinfo{person}{Pere Barlet{-}Ros}, \bibinfo{person}{Eduard Alarc{\'{o}}n},
  \bibinfo{person}{Marc Sol{\'{e}}}, \bibinfo{person}{Victor
  Munt{\'{e}}s{-}Mulero}, \bibinfo{person}{David Meyer},
  \bibinfo{person}{Sharon Barkai}, \bibinfo{person}{Mike~J. Hibbett},
  \bibinfo{person}{Giovani Estrada}, \bibinfo{person}{Khaldun Maruf},
  \bibinfo{person}{Florin Coras}, \bibinfo{person}{Vina Ermagan},
  \bibinfo{person}{Hugo Latapie}, \bibinfo{person}{Chris Cassar},
  \bibinfo{person}{John Evans}, \bibinfo{person}{Fabio Maino},
  \bibinfo{person}{Jean~C. Walrand}, {and} \bibinfo{person}{Albert Cabellos}.}
  \bibinfo{year}{2017}\natexlab{}.
\newblock \showarticletitle{Knowledge-Defined Networking}.
\newblock \bibinfo{journal}{\emph{Computer Communication Review}}
  \bibinfo{volume}{47}, \bibinfo{number}{3} (\bibinfo{year}{2017}),
  \bibinfo{pages}{2--10}.
\newblock


\bibitem[\protect\citeauthoryear{Miao, Zeng, Kim, Lee, and Yu}{Miao
  et~al\mbox{.}}{2017}]%
        {MiaoZKLY17}
\bibfield{author}{\bibinfo{person}{Rui Miao}, \bibinfo{person}{Hongyi Zeng},
  \bibinfo{person}{Changhoon Kim}, \bibinfo{person}{Jeongkeun Lee}, {and}
  \bibinfo{person}{Minlan Yu}.} \bibinfo{year}{2017}\natexlab{}.
\newblock \showarticletitle{SilkRoad: Making Stateful Layer-4 Load Balancing
  Fast and Cheap Using Switching ASICs}. In \bibinfo{booktitle}{\emph{Proc. of
  ACM {SIGCOMM}}}. \bibinfo{pages}{15--28}.
\newblock


\bibitem[\protect\citeauthoryear{Miguel, Albericio, Moshovos, and
  Jerger}{Miguel et~al\mbox{.}}{2015}]%
        {MiguelAMJ15}
\bibfield{author}{\bibinfo{person}{Joshua~San Miguel}, \bibinfo{person}{Jorge
  Albericio}, \bibinfo{person}{Andreas Moshovos}, {and}
  \bibinfo{person}{Natalie D.~Enright Jerger}.}
  \bibinfo{year}{2015}\natexlab{}.
\newblock \showarticletitle{Doppelg{\"{a}}nger: a cache for approximate
  computing}. In \bibinfo{booktitle}{\emph{Proc. of the 48th Int. Symp. on
  Microarchitecture ({MICRO})}}. \bibinfo{pages}{50--61}.
\newblock


\bibitem[\protect\citeauthoryear{Neugebauer, Antichi, Zazo, Audzevich,
  L\'{o}pez-Buedo, and Moore}{Neugebauer et~al\mbox{.}}{2018}]%
        {pcie}
\bibfield{author}{\bibinfo{person}{Rolf Neugebauer}, \bibinfo{person}{Gianni
  Antichi}, \bibinfo{person}{Jos\'{e}~Fernando Zazo}, \bibinfo{person}{Yury
  Audzevich}, \bibinfo{person}{Sergio L\'{o}pez-Buedo}, {and}
  \bibinfo{person}{Andrew~W. Moore}.} \bibinfo{year}{2018}\natexlab{}.
\newblock \showarticletitle{Understanding {PCIe} Performance for End Host
  Networking}. In \bibinfo{booktitle}{\emph{Proc. of {ACM} {SIGCOMM}}}.
  \bibinfo{pages}{327–341}.
\newblock


\bibitem[\protect\citeauthoryear{Pacheco, Exposito, Gineste, Baudoin, and
  Aguilar}{Pacheco et~al\mbox{.}}{2019}]%
        {PachecoEGBA19}
\bibfield{author}{\bibinfo{person}{Fannia Pacheco}, \bibinfo{person}{Ernesto
  Exposito}, \bibinfo{person}{Mathieu Gineste}, \bibinfo{person}{C{\'{e}}dric
  Baudoin}, {and} \bibinfo{person}{Jos{\'{e}} Aguilar}.}
  \bibinfo{year}{2019}\natexlab{}.
\newblock \showarticletitle{Towards the Deployment of Machine Learning
  Solutions in Network Traffic Classification: {A} Systematic Survey}.
\newblock \bibinfo{journal}{\emph{{IEEE} Commun. Surv. Tutorials}}
  \bibinfo{volume}{21}, \bibinfo{number}{2} (\bibinfo{year}{2019}),
  \bibinfo{pages}{1988--2014}.
\newblock


\bibitem[\protect\citeauthoryear{Pandey, Broder, Chierichetti, Josifovski,
  Kumar, and Vassilvitskii}{Pandey et~al\mbox{.}}{2009}]%
        {PandeyBCJKV09}
\bibfield{author}{\bibinfo{person}{Sandeep Pandey}, \bibinfo{person}{Andrei~Z.
  Broder}, \bibinfo{person}{Flavio Chierichetti}, \bibinfo{person}{Vanja
  Josifovski}, \bibinfo{person}{Ravi Kumar}, {and} \bibinfo{person}{Sergei
  Vassilvitskii}.} \bibinfo{year}{2009}\natexlab{}.
\newblock \showarticletitle{Nearest-neighbor caching for content-match
  applications}. In \bibinfo{booktitle}{\emph{Proc. of {WWW} Conf.}}
  \bibinfo{pages}{441--450}.
\newblock


\bibitem[\protect\citeauthoryear{Reuther, Michaleas, Jones, Gadepally, Samsi,
  and Kepner}{Reuther et~al\mbox{.}}{2019}]%
        {HPECsurvey19}
\bibfield{author}{\bibinfo{person}{Albert Reuther}, \bibinfo{person}{Peter
  Michaleas}, \bibinfo{person}{Michael Jones}, \bibinfo{person}{Vijay
  Gadepally}, \bibinfo{person}{Siddharth Samsi}, {and} \bibinfo{person}{Jeremy
  Kepner}.} \bibinfo{year}{2019}\natexlab{}.
\newblock \showarticletitle{Survey and Benchmarking of Machine Learning
  Accelerators}. In \bibinfo{booktitle}{\emph{IEEE High Perf. Extreme Comp.
  (HPEC)}}.
\newblock


\bibitem[\protect\citeauthoryear{Rizzo}{Rizzo}{2012}]%
        {netmap}
\bibfield{author}{\bibinfo{person}{Luigi Rizzo}.}
  \bibinfo{year}{2012}\natexlab{}.
\newblock \showarticletitle{netmap: A Novel Framework for Fast Packet I/O}. In
  \bibinfo{booktitle}{\emph{In Proc of {USENIX} {ATC}}}.
  \bibinfo{pages}{101--112}.
\newblock


\bibitem[\protect\citeauthoryear{Sanvito, Siracusano, and Bifulco}{Sanvito
  et~al\mbox{.}}{2018}]%
        {SanvitoSB18}
\bibfield{author}{\bibinfo{person}{Davide Sanvito}, \bibinfo{person}{Giuseppe
  Siracusano}, {and} \bibinfo{person}{Roberto Bifulco}.}
  \bibinfo{year}{2018}\natexlab{}.
\newblock \showarticletitle{Can the Network be the {AI} Accelerator?}. In
  \bibinfo{booktitle}{\emph{NetCompute Workshop at ACM SIGCOMM}}.
\newblock


\bibitem[\protect\citeauthoryear{Sapio, Abdelaziz, Aldilaijan, Canini, and
  Kalnis}{Sapio et~al\mbox{.}}{2017}]%
        {SapioAACK17}
\bibfield{author}{\bibinfo{person}{Amedeo Sapio}, \bibinfo{person}{Ibrahim
  Abdelaziz}, \bibinfo{person}{Abdulla Aldilaijan}, \bibinfo{person}{Marco
  Canini}, {and} \bibinfo{person}{Panos Kalnis}.}
  \bibinfo{year}{2017}\natexlab{}.
\newblock \showarticletitle{In-Network Computation is a Dumb Idea Whose Time
  Has Come}. In \bibinfo{booktitle}{\emph{{ACM} {HotNet} Workshop}}.
  \bibinfo{pages}{150--156}.
\newblock


\bibitem[\protect\citeauthoryear{Siracusano and Bifulco}{Siracusano and
  Bifulco}{2018}]%
        {innetworkNN18}
\bibfield{author}{\bibinfo{person}{Giuseppe Siracusano} {and}
  \bibinfo{person}{Roberto Bifulco}.} \bibinfo{year}{2018}\natexlab{}.
\newblock \showarticletitle{In-network Neural Networks}.
\newblock \bibinfo{journal}{\emph{CoRR}}  \bibinfo{volume}{abs/1801.05731}
  (\bibinfo{year}{2018}).
\newblock
\showeprint[arxiv]{1801.05731}
\urldef\tempurl%
\url{http://arxiv.org/abs/1801.05731}
\showURL{%
\tempurl}


\bibitem[\protect\citeauthoryear{Siracusano, Galea, Sanvito, Malekzadeh,
  Haddadi, Antichi, and Bifulco}{Siracusano et~al\mbox{.}}{2020}]%
        {abs-2009-02353}
\bibfield{author}{\bibinfo{person}{Giuseppe Siracusano},
  \bibinfo{person}{Salvator Galea}, \bibinfo{person}{Davide Sanvito},
  \bibinfo{person}{Mohammad Malekzadeh}, \bibinfo{person}{Hamed Haddadi},
  \bibinfo{person}{Gianni Antichi}, {and} \bibinfo{person}{Roberto Bifulco}.}
  \bibinfo{year}{2020}\natexlab{}.
\newblock \showarticletitle{Running Neural Networks on the {NIC}}.
\newblock \bibinfo{journal}{\emph{CoRR}}  \bibinfo{volume}{abs/2009.02353}
  (\bibinfo{year}{2020}).
\newblock
\showeprint[arxiv]{2009.02353}
\urldef\tempurl%
\url{https://arxiv.org/abs/2009.02353}
\showURL{%
\tempurl}


\bibitem[\protect\citeauthoryear{Swamy, Rucker, Shahbaz, and Olukotun}{Swamy
  et~al\mbox{.}}{2020}]%
        {swamy2020taurus}
\bibfield{author}{\bibinfo{person}{Tushar Swamy}, \bibinfo{person}{Alexander
  Rucker}, \bibinfo{person}{Muhammad Shahbaz}, {and} \bibinfo{person}{Kunle
  Olukotun}.} \bibinfo{year}{2020}\natexlab{}.
\newblock \bibinfo{title}{Taurus: An Intelligent Data Plane}.
\newblock
\newblock
\showeprint[arxiv]{cs.NI/2002.08987}


\bibitem[\protect\citeauthoryear{Van, Tran, Souihi, and Abdelhamid}{Van
  et~al\mbox{.}}{2018}]%
        {troubleshooting}
\bibfield{author}{\bibinfo{person}{T Van}, \bibinfo{person}{Hai~Anh Tran},
  \bibinfo{person}{Sami Souihi}, {and} \bibinfo{person}{Mellouk Abdelhamid}.}
  \bibinfo{year}{2018}\natexlab{}.
\newblock \showarticletitle{Network troubleshooting: Survey, Taxonomy and
  Challenges.}. In \bibinfo{booktitle}{\emph{Proc. of International Conference
  on Smart Communications in Network Technologies (SaCoNeT)}}.
\newblock


\bibitem[\protect\citeauthoryear{Vinod, Wang, and Bal}{Vinod
  et~al\mbox{.}}{2020}]%
        {nigade2020clownfish}
\bibfield{author}{\bibinfo{person}{Nigade Vinod}, \bibinfo{person}{Lin Wang},
  {and} \bibinfo{person}{Henri Bal}.} \bibinfo{year}{2020}\natexlab{}.
\newblock \showarticletitle{{Clownfish}: Edge and Cloud Symbiosis for Video
  Stream Analytics}. In \bibinfo{booktitle}{\emph{{ACM/IEEE} Symposium on Edge
  Computing ({SEC})}}.
\newblock


\bibitem[\protect\citeauthoryear{Wang, Gao, Zhang, Wang, Chen, Ng, Ooi, Shao,
  and Reyad}{Wang et~al\mbox{.}}{2018}]%
        {WangWGZCNOS18}
\bibfield{author}{\bibinfo{person}{Wei Wang}, \bibinfo{person}{Jinyang Gao},
  \bibinfo{person}{Meihui Zhang}, \bibinfo{person}{Sheng Wang},
  \bibinfo{person}{Gang Chen}, \bibinfo{person}{Teck~Khim Ng},
  \bibinfo{person}{Beng~Chin Ooi}, \bibinfo{person}{Jie Shao}, {and}
  \bibinfo{person}{Moaz Reyad}.} \bibinfo{year}{2018}\natexlab{}.
\newblock \showarticletitle{Rafiki: Machine Learning as an Analytics Service
  System}.
\newblock \bibinfo{journal}{\emph{Proc. {VLDB} Endow.}} \bibinfo{volume}{12},
  \bibinfo{number}{2} (\bibinfo{year}{2018}), \bibinfo{pages}{128--140}.
\newblock


\bibitem[\protect\citeauthoryear{Woo and Park}{Woo and Park}{2012}]%
        {symmtoeplitz}
\bibfield{author}{\bibinfo{person}{Shinae Woo} {and} \bibinfo{person}{KyoungSoo
  Park}.} \bibinfo{year}{2012}\natexlab{}.
\newblock \bibinfo{title}{Scalable TCP Session Monitoring with Symmetric
  Receive-side Scaling}.
\newblock
  \bibinfo{howpublished}{\url{http://an.kaist.ac.kr/~shinae/paper/2012-srss.pdf}}.
\newblock


\bibitem[\protect\citeauthoryear{Xiong and Zilberman}{Xiong and
  Zilberman}{2019}]%
        {XiongZ19}
\bibfield{author}{\bibinfo{person}{Zhaoqi Xiong} {and} \bibinfo{person}{Noa
  Zilberman}.} \bibinfo{year}{2019}\natexlab{}.
\newblock \showarticletitle{Do Switches Dream of Machine Learning?: Toward
  In-Network Classification}. In \bibinfo{booktitle}{\emph{18th {ACM} Workshop
  HotNets}}.
\newblock


\bibitem[\protect\citeauthoryear{Xu, Tang, Meng, Zhang, Wang, Liu, and Yang}{Xu
  et~al\mbox{.}}{2018}]%
        {XuTMZWLY18}
\bibfield{author}{\bibinfo{person}{Zhiyuan Xu}, \bibinfo{person}{Jian Tang},
  \bibinfo{person}{Jingsong Meng}, \bibinfo{person}{Weiyi Zhang},
  \bibinfo{person}{Yanzhi Wang}, \bibinfo{person}{Chi~Harold Liu}, {and}
  \bibinfo{person}{Dejun Yang}.} \bibinfo{year}{2018}\natexlab{}.
\newblock \showarticletitle{Experience-driven Networking: {A} Deep
  Reinforcement Learning based Approach}. In \bibinfo{booktitle}{\emph{{IEEE}
  {Infocom}}}. \bibinfo{pages}{1871--1879}.
\newblock


\bibitem[\protect\citeauthoryear{Zhang, Yu, Wang, and Yan}{Zhang
  et~al\mbox{.}}{2019}]%
        {ZhangYWY19}
\bibfield{author}{\bibinfo{person}{Chengliang Zhang}, \bibinfo{person}{Minchen
  Yu}, \bibinfo{person}{Wei Wang}, {and} \bibinfo{person}{Feng Yan}.}
  \bibinfo{year}{2019}\natexlab{}.
\newblock \showarticletitle{MArk: Exploiting Cloud Services for Cost-Effective,
  SLO-Aware Machine Learning Inference Serving}. In
  \bibinfo{booktitle}{\emph{{USENIX} {ATC}}}. \bibinfo{pages}{1049--1062}.
\newblock


\end{thebibliography}

\end{document}